\documentclass[fleqn,usenatbib,useAMS,twocolumn]{mnras}

\usepackage[T1]{fontenc}
\usepackage{ae,aecompl}

\usepackage{newtxtext,newtxmath}

\usepackage{graphicx}
\usepackage{subfig}
\usepackage{amsmath}
\usepackage{xfrac}
\usepackage{physics}

\usepackage{scalerel}
\usepackage{tikz}
\usetikzlibrary{svg.path}

\definecolor{orcidlogocol}{HTML}{A6CE39}
\tikzset{
  orcidlogo/.pic={
    \fill[orcidlogocol] svg{M256,128c0,70.7-57.3,128-128,128C57.3,256,0,198.7,0,128C0,57.3,57.3,0,128,0C198.7,0,256,57.3,256,128z};
    \fill[white] svg{M86.3,186.2H70.9V79.1h15.4v48.4V186.2z}
                 svg{M108.9,79.1h41.6c39.6,0,57,28.3,57,53.6c0,27.5-21.5,53.6-56.8,53.6h-41.8V79.1z M124.3,172.4h24.5c34.9,0,42.9-26.5,42.9-39.7c0-21.5-13.7-39.7-43.7-39.7h-23.7V172.4z}
                 svg{M88.7,56.8c0,5.5-4.5,10.1-10.1,10.1c-5.6,0-10.1-4.6-10.1-10.1c0-5.6,4.5-10.1,10.1-10.1C84.2,46.7,88.7,51.3,88.7,56.8z};
  }
}

\newcommand\orcidicon[1]{\href{https://orcid.org/#1}{\mbox{\scalerel*{
\begin{tikzpicture}[yscale=-1,transform shape]
\pic{orcidlogo};
\end{tikzpicture}
}{|}}}}

\usepackage{cleveref}

\newcommand{\pd}[2]{\partial_{#1}{#2}}

\renewcommand{\vec}[1]{\mathbf{#1}}

\newcommand{\gvec}[1]{\boldsymbol{#1}}

\newcommand{\parallelsum}{\mathrel{/\mkern-5mu/}}

\newcommand{\change}[1]{#1}

\usepackage{soul}

\newcommand{\remove}[1]{}

\def\O{{\mathcal O}}

\def\S{{\mathcal S}}

\def\eps{{\epsilon}}
\def\veps{{\varepsilon}}
\def\om{{\omega}}
\def\Om{{\Omega}}

\def\varp{{\varphi}}

\title[Local magneto-shear instability]{Local magneto-shear instability in Newtonian gravity}

\author[T. Celora, I. Hawke, N. Andersson and G.L. Comer]{Thomas Celora \orcidicon{0000-0002-6515-3644},$^{1}$\thanks{E-mail: \href{mailto:T.Celora@soton.ac.uk}{T.Celora@soton.ac.uk}} Ian Hawke \orcidicon{0000-0003-4805-0309},$^1$ Nils Andersson \orcidicon{0000-0001-8550-3843}$^1$ and Gregory L. Comer \orcidicon{0000-0002-7937-1490}$^{2}$\thanks{E-mail: \href{mailto:comergl@slu.edu}{comergl@slu.edu}} \\
$^1$ School of Mathematics and STAG Research Centre, University of Southampton, Southampton SO17 1BJ, UK \\
$^2$  Department of Physics, Saint Louis University,
St Louis, MO 63156-0907, USA}

\pubyear{2023}

\begin{document}
\maketitle

\date{} 

\begin{abstract}
The magneto-rotational instability (MRI)---which is due to an interplay between a sheared background and the magnetic field---is commonly considered a key ingredient for developing and sustaining turbulence in the outer envelope of binary neutron star merger remnants. 
To assess whether (or not) the instability is active and resolved, criteria originally derived in the accretion disk literature---thus exploiting the symmetries of such systems---are often used. 
In this paper we discuss the magneto-shear instability as a truly local phenomenon, relaxing common symmetry assumptions on the background on top of which the instability grows. 
This makes the discussion well-suited for highly dynamical environments such as binary mergers. 
We find that---although this is somewhat hidden in the usual derivation of the MRI dispersion relation---the instability crucially depends on the assumed symmetries. 
Relaxing the symmetry assumptions on the background we find that the role of the magnetic field is significantly diminished, as it affects the modes' growth but does not drive it.
\remove{This suggests that we should not expect the standard instability criteria to provide a faithful indication/diagnostic of what ``is actually going on'' in mergers.}
We conclude by making contact with a suitable filtering operation, as this is key to separating background and fluctuations in highly dynamical systems. 
\end{abstract}


\maketitle

\section{Introduction}

The magneto-rotational instability was discovered by Balbus and Hawley in the early 1990s~\citep{BalbusHawley1,BalbusHawley2,BalbusHawleyRev} (linking to earlier ideas from, for example, \citet{Chandrasekar1960} and \citet{velikhov1959}). Due to the fast instability  growth rate, this mechanism is considered the most promising candidate for developing/sustaining magneto-hydrodynamic turbulence in accretion disks as well as  explaining enhanced angular momentum transfer \citep{BalbusHawleyRev,Shakura+1973}. 
The instability is due to an interplay between a weak magnetic field and a sheared background flow. With few exceptions (see for example \citet{MahajanKrishan2008} and \citet{Shakura2022}) and due to its ``local'' nature, the magneto-rotational instability is commonly  discussed in the so-called ``shearing box approximation'' \citep{GoldreichLynden-Bell65,Hill1878}. That is, the instability is established in  a frame that corotates with a fiducial point in the mid-plane of the undisturbed disk \citep[see also][]{GoodmanXu94}. 
This is convenient for analytical studies as well as numerical analysis since local simulations can reach much higher resolutions than global ones \citep[see, for example,][and references therein]{1995ApJ...440..742H,ZierVolker22}. 
\change{Shearing box simulations confirmed the predictions of the linear theory, and allowed studies of non-linear features, such as the formation of the so-called channel modes, eventually leading to a turbulent regime as these are destroyed by a parasitic instability \citep{GoodmanXu94}. As turbulence decays, the magneto-rotational instability may be revived and the process can start again in cycles. Most importantly, while it has recently become possible to perform fully kinetic simulations \citep[see, e.g. ][]{Hoshino, Inchingolo_2018}, shearing box simulations are necessary to explore the parameter space in a variety of different setups \citep[][]{Sharma_2006,Kempski2019,GuiletHighPm,held2022}. }

Although originally discussed in the context of accretion disks, the magneto-rotational instability is also thought to play a role in neutron-star mergers \citep{Duez+2006,Siegel2013,Palenzuela2022,Margalit2022,Hayashi22_1s,Kiuchi2022_1s}, especially for sustaining a magneto-turbulent state in the outer envelope of the remnant, where the Kelvin-Helmholtz instability is less significant or, indeed, not active \citep{Kiuchi:2017zzg}. 
To assess whether or not the magneto-rotational instability is  active and resolved in merger simulations, criteria discussed/established in the context of accretion disks \citep{Hawley:2011tq,Hawley:2013lga,shibataNR} are often used. 
However, because binary neutron star mergers are highly dynamical environments, framing a discussion of the magneto-rotational instability using criteria that exploits restrictive symmetry conditions might be misleading.
Motivated by this, we aim to explore the impact of relaxing common assumptions---well-motivated in the accretion disks scenario, like an \change{axisymmetric} and circular background flow, but less so for mergers---on the magneto-shear instability.

The paper is laid out as follows: we start in \cref{sec:WKBlocalbox} by introducing the WKB-type expansion together with the concept of fast vs. slow background gradients, as needed to derive dispersion relations associated with perturbations propagating on top of a non-homogeneous background flow.
In \cref{sec:SlowlyEvolving} we discuss how we can, in highly dynamical environments such as mergers, still refer to an unspecified background---with no symmetries stated from the outset---and consider perturbations rapidly evolving on top of that. 
We continue in \cref{sec:NonInertialEq} by deriving the non-inertial induction equation, and making contact with the concept of a local frame. 
We do so as we will study the instability from the perspective of an observer moving around with the background flow, which cannot be expected to be inertial on a general basis. 
We discuss our main findings in sections \ref{sec:Back2Hydro} and \ref{sec:MSinstaLocal}. 
These are based on the dispersion relations, which are derived in detail in \cref{app:HydroDispRel,app:MHDdispRel} in order to be able to focus on the physics in the main body of the paper. 
In particular, we discuss the magneto-shear instability paying careful attention to the differences introduced by the addition of a magnetic field.
We draw our conclusions, and make contact with the magneto-rotational instability (and the associated criteria) in \cref{sec:Conclusions}. 

\section{Background gradients and plane-wave expansion}\label{sec:WKBlocalbox}

Let us begin by noting that the magneto-rotational instability is, in some sense, a ``global instability analyzed with local tools''. 
The local nature is  evident since the instability is established by means of a dispersion relation (hence involves a plane-wave expansion and, by assumption, a short-wavelength approximation). 
At the same time, one may appreciate the ``global nature'' of the instability by recalling the key aspects of the instability: the addition of a weak magnetic field turns  axisymmetric modes (which would otherwise be hydrodynamically stable) unstable.
The global axisymmetry of the background, then, plays a crucial role as the relevant hydrodynamic stability criterion---the Rayleigh criterion \citep{Rayleigh1917}---applies to axisymmetric modes only.
Although the standard derivation of the instability does not highlight this subtlety, this  aspect becomes apparent if we formulate the problem using a co-rotating local frame (cf. the discussion in \cref{app:MRIlocal,app:RayleighCloser}).

With these points in mind, let us  spell out how we intend to discuss the magneto-shear instability without referring to a given axisymmetric and circular background. Consistent with the shearing box idea \citep{GoldreichLynden-Bell65,Hill1878}, the strategy is to zoom in on a small region of fluid---small enough for the analysis to be local but large enough to allow for a meaningful hydrodynamic description. 
We then set up a local Cartesian frame co-moving with the background flow---so that the background velocity vanishes at the origin of the local box. 
As this frame moves with the flow---and hence cannot be expected to be inertial---we need to consider the (at this point Newtonian) ideal magneto-hydrodynamics equations in a non-inertial frame. 
This step is commonly left out of the discussion, so we first of all have to fill this gap. The non-inertial equations will then be perturbed---retaining gradients in the background quantities as explained below---and a local WKB-type dispersion relation will be derived and studied.  
This way we can account for the effects of a sheared background and its interplay with the magnetic field in a general setting. 

Strictly speaking, the plane-wave expansion only makes sense for a homogeneous background---that is, the plane-wave amplitude is assumed to vary on the same scales as the background. 
At the same time, we know that a sheared background is key to the magneto-rotational instability. 
Therefore, given any quantity/field $a$, we first write it as a sum of background plus perturbations
\begin{equation}
    a = A + \delta A \;,
\end{equation}
and then introduce a WKB-type expansion of the form \citep{ThorneBlandford,Anile}
\begin{equation}\label{eq:WKBexp}
    \delta A = \left(\sum_{q=0} \epsilon^q \bar A_q \right)e^{i\theta/\eps}\; \, \bar\delta ,
\end{equation}
with book-keeping parameters $\bar\delta$ and $\eps$ \citep[see also][]{Palapanidis}. 
The former $\left(\bar \delta\right)$ is introduced to measure the relative magnitude of background vs. perturbations, while the latter ($\epsilon$) is given by $\eps \approx \lambda / L$ where $\lambda$ is the typical wavelength of the waves and $L$ is the typical lengthscale over which the wave amplitude, polarization and wavelength vary.
Having split the perturbations into amplitude and phase, we follow the standard convention \citep{GravitationMTW} and stick all ``post-geometric optics'' corrections into the amplitude $\bar A_q$. 
With this Ansatz, the background equations are obtained by collecting all terms of order $\O(\bar\delta^0, \eps^0)$, while the perturbation equations are obtained collecting terms of order $\O(\bar\delta^1, \eps^0)$. 
Terms of higher order in $\eps$ correspond to post-geometric optics, while those of higher order in $\bar \delta$ represent non-linear perturbations. 

Along with this WKB-type Ansatz, we need to introduce the concept of fast and slowly varying quantities. Given a specific choice of coordinates, a quantity is slow in the variable $x$ if $A = A(X) $ where $X = \eps x$ while it is fast if $A = A(x)$. 
Deciding which quantities are fast or slow corresponds to specifying (in a qualitative manner) the background configuration. 
As an illustration, consider the simple toy problem 
\begin{equation}
    a (\partial_x b + \partial_x c) =0 \;,
\end{equation}
together with the Ansatz from \cref{eq:WKBexp}. 
Let us first assume that both $B$ and $C$ are fast, so that $\partial_x B \approx \O(\bar\delta^0,\eps^0)$ and similarly for $C$. The background equation is then
\begin{equation}
    A\left(\partial_x B + \partial_x C\right) = 0 \;.
\end{equation}
If we  instead assume that, say, $B$ is fast while $C$ is slow, then $\partial_x B \approx \O(\bar\delta^0, \eps^0)$ while $\partial_x C \approx \O(\bar\delta^0,\eps)$ and the background equation becomes
\begin{equation}
    A \partial_xB =0\;.
\end{equation}
Clearly, the two problems are different already at the background level.

Let us now turn to the linear perturbations. 
Because we have explicitly introduced the book-keeping parameter $\eps$ in \cref{eq:WKBexp}, we take all amplitude terms as well as the phase to be slowly varying. 
Then, to order $\O(\bar\delta,\eps^0)$ we have
\begin{equation}
    \left(A + \bar\delta \bar A_0 e^{i\theta/\eps}\right) \partial_x  \left[B + \bar\delta \bar B_0 e^{i\theta/\eps} + C + \bar\delta \bar C_0 e^{i\theta/\eps}\right] =0 \;.
\end{equation}
Assuming again that the background quantity $B$ is fast, while $C$ is slow, the perturbation equation becomes 
\begin{equation}
    \bar A_0 \left(\partial_x B\right) e^{i\theta/\eps} + A\left(\bar B_0 \partial_x e^{i\theta/\eps} + \bar C_0 \partial_x e^{i\theta/\eps}\right) =0 \;.
\end{equation}
Next, Taylor expanding the phase---which is slowly varying---we get
\begin{equation}
    \frac{\theta(x)}{\eps} \approx \frac{\theta(0)}{\eps} + \pdv{\theta}{X}\Big|_{X=0}x + \dots = \theta(0)/\eps + k_x x + \O(\eps) \;,
\end{equation}
where we define the wave-vector $k_x = \partial\theta / \partial X$ from the first order term in the expansion,  while the overall constant can be neglected.
Then, introducing an analogous expansion for the fast background gradients 
$\partial_x B (x)= \partial_x B(0) + \O(\eps)$ we end up with
\begin{equation}
    \bar A_0 (\partial_x B) + A\left(ik_x \bar B_0 + i k_x \bar C_0\right) =0 \;,
\end{equation}
where both $\partial_x B$ and $A$ are evaluated at a point (conveniently chosen as the origin of the coordinate system).
Therefore, if all background quantities are ``slow'', we get back the dispersion relation we would have obtained ignoring all background gradients. This is quite intuitive. 
However, the strategy also allows us to account for the impact that ``fast'' background gradients have on the dispersion relation.
In short, as long as these terms are treated as constants, we may retain them and work out a dispersion relation in the usual way. 

\section{The slowly evolving background}\label{sec:SlowlyEvolving}

The starting point for any hydrodynamic perturbation analysis is the choice/identification of a stationary background flow configuration, which is then perturbed in order to establish stability (or not). 
Here, we want to frame the analysis of the magneto-shear instability without considering a specific background configuration with constraining symmetries stated from the outset.  Nonetheless, we need to clarify how we can refer to a suitable ``background'' in highly dynamical environments like binary neutron star mergers. 
We will first analyse the problem analytically, and return to discuss the link to numerical simulations in \cref{sec:Conclusions}.
Given real numerical simulation data, this discussion will inevitably involve some kind of filtering operation. Anticipating that this can be done in a meaningful way, we consider perturbations evolving rapidly  with respect to the evolution time-scale of an unspecified ``background'' flow. 

To make this statement more precise, let us consider the inertial ideal MHD equations and introduce reference values for each quantity (indicated with an ``$r$'' subscript) such as $\rho = \rho_r \tilde \rho$. 
We introduce the (dimensionless)
Strouhal, Mach, Froude and magnetic interaction numbers as
\begin{equation}
    \veps_\text{St} = \frac{l_r}{t_rv_r} \;,  \quad \veps_\text{Ma} = \frac{v_r}{c_r}\;, \quad \veps_\text{Fr} = \frac{v_r}{\sqrt{\Phi_r}} \;, \quad \veps_{B} = \frac{B_r^2}{\mu_0\rho_r v_r^2} \;,
\end{equation}
where $l_r,\,t_r,\,v_r$ are characteristic lengthscale, timescale and velocity (respectively) while $B_r,\,\Phi_r,\,\rho_r$ are reference values for the magnetic field, gravitational potential and density and $c_r$ is the (adiabatic) speed of sound. 
This way, the non-dimensional inertial ideal MHD equations read (now dropping the tildes for notational clarity)
\begin{subequations}
\begin{equation}
    \veps_\text{St} \,\pd{t}{\rho} = - \rho \nabla_i v^i - v^i \nabla_i \rho \;,
\end{equation}
\begin{equation}
    \veps_\text{St} \,\pd{t}{B^i} = - v^j \nabla_j B^i + B^j\nabla_j v^i - B^i \nabla_jv^j \;,
\end{equation}
\begin{multline}
    \veps_\text{St}\, \pd{t}{v^i} = - v^j \nabla_j v^i -\frac{1}{\eps_{\text{Ma}}^2} \frac{1}{\rho}\nabla^i\rho -\frac{1}{\eps_{\text{Fr}}^2} \nabla^i \Phi - \\ \eps_{B}\frac{1}{\rho} \left[  B^j  \nabla_j B^i -\nabla^i \left(\frac{B^2}{2}\right) \right] \;.
\end{multline}
\end{subequations}
From this we see that a generic flow configuration can be considered slowly evolving (in time) as long as the corresponding Strouhal number is small. 
In practice, given a characteristic lengthscale $l_r$ and velocity $v_r$ of a generic flow, we consider disturbances evolving on timescales $t_r$ such that $\eps_{\text{St}}\ll 1$---over which the background can be effectively taken as stationary.
In turn, this  determines the time-scales over which we expect the following results to be reliable.

\subsection{Velocity gradient decomposition}\label{subsec:vel_gradients}
In the following we will consider the impact that gradients in the background flow velocity have on the time evolution of perturbations.
It is then convenient to introduce the standard decomposition of the velocity gradient into expansion, shear and vorticity. That is, 
\begin{equation}
     \nabla_i v_j =  \frac{1}{3}\theta g_{ij} + \sigma_{ij} + \om_{ij} \;,
\end{equation}
where\footnote{We follow the standard notation and use (straight) round brackets to indicate (anti-)symmetrization of the indices enclosed.}
\begin{subequations}\label{eq:velGradDec}
\begin{align}
    &\theta = \nabla_i v^i \;, \\
    &\sigma_{ij} =\nabla_{(i}v_{j)} - \frac{1}{3}\theta g_{ij}= \frac{1}{2}\left(\nabla_{i}v_{j}+ \nabla_j v_i\right) - \frac{1}{3}\theta g_{ij}\;, \\
    &\omega_{ij}=  \nabla_{[i}v_{j]} = \frac{1}{2}\left(\nabla_{i}v_{j}- \nabla_j v_i\right)  \;.
\end{align}
\end{subequations}
In order to bring out the magneto-shear nature of the instability, we will consider the impact of having a background with non-negligible shear and vorticity separately. 
We will, however, not consider the impact of a background expansion rate as exact non-linear results are sufficient to predict this.  In fact, due to the Alfvén theorem, we  know that the magnetic intensity must grow in a (ideal magneto-)fluid undergoing compression as the field lines are  squeezed together.  Similarly, the field will get weaker in an expanding fluid. In essence, 
we expect---and have verified explicitly---this non-linear prediction to emerge in the  analysis as a generic ``instability''. The background magnetic field cannot grow in time as it is assumed to be slowly evolving by construction, so the required growth must be represented by perturbations.

Before we move on to derive the non-inertial equations, it is useful to take a brief detour and consider a realization of a flow with only non-negligible shear.
Because we are interested in flows that are slowly evolving we can start by assuming that $v^i = v^i(\eps t,x,y,z)$ and suppress the time dependence in the following.
We then take the velocity vector as mainly two-dimensional, specifically in the $x-y$ plane of a set of local Cartesian coordinates
\begin{equation}
    \vec v = v^x( x,y,z) \hat x+ v^y(x, y, z) \hat y + \O(\eps) \;,
\end{equation}
where, in order to make sure the expansion is small, we take $\pd{x}{v^x} = - \pd{y}{v^y} + \O(\eps)$.
The shear matrix  is then given by
\begin{equation}
    \gvec{\sigma} = \begin{pmatrix}
    \pd{x}{v^x} & \frac{1}{2}\left(\pd{x}{v^y} + \pd{y}{v^x}\right) & \frac{1}{2}\pd{z}{v^x} \\
    \frac{1}{2}\left(\pd{x}{v^y} + \pd{y}{v^x}\right) & - \pd{x}{v^x}  & \frac{1}{2}\pd{z}{v^y} \\
    \frac{1}{2}\pd{z}{v^x} & \frac{1}{2}\pd{z}{v^y} & 0
    \end{pmatrix} + \O(\eps) \;,
\end{equation}
while the curl of $\vec v$ becomes
\begin{equation}
    \nabla \times \vec v = -\left(\pd{z}{v^y}\right) \hat x + \left(\pd{z}{v^x} \right)\hat y  + \left( \pd{x}{v^y} - \pd{y}{v^x}\right) \hat z \;.
\end{equation}
This has to be of $\O(\eps)$ for background flows with only non-negligibile shear, in which case
\begin{equation}
    \pd{z}{v^x} = \pd{z}{v^y} =  \O(\eps) \;, \qquad \pd{y}{v^x} = \pd{x}{v^y} + \O(\eps), 
\end{equation}
and as result the determinant of the shear matrix vanishes (more precisely, is of order $\O(\eps)$).
This is equivalent to saying that two eigenvalues of the shear matrix are opposite and the third is zero (to order $\O(\eps)$). 
In essence, a mainly two-dimensional flow with negligible expansion and vorticity is characterized by a shear matrix with one zero eigenvalue, and hence a vanishing determinant.
This will turn out to be a useful observation when we discuss the dispersion relations in \cref{sec:Back2Hydro,sec:MSinstaLocal}.
We also note that having negligible expansion (although relevant for the present analysis) is not strictly necessary to the argument. 

\section{Non-inertial equations and the local frame}\label{sec:NonInertialEq}
The ideal magneto-hydrodynamics equations above hold in an inertial frame. 
However, as an observer locally co-moving with the fluid cannot  (in general) be expected to be inertial, we need to consider the equations according to a non-inertial observer. 
As such, let us now derive the Newtonian induction equation in a general (non-inertial) frame starting from the Maxwell equations formulated in covariant form. 
This is the natural starting point as electromagnetism is inherently a relativistic theory.
Moreover, we stress that the electric and magnetic field are observer dependent quantities, and the standard textbook form of the Maxwell equations assume an inertial observer/frame from the outset. 

We then take as our starting point the covariant Maxwell equations (with indices $a,b,\ldots$ representing space-time components, in contrast to the $i,j,k,\ldots$ components from before)
\begin{equation}
    \nabla_aF^{ba} = \mu_0 j^b \;, \qquad \nabla_{[a}F_{bc]} = 0 \;,
\end{equation}
where $F^{ab}$ is the Faraday tensor while $j^b$ is the four-current. 
From these relations we derive the non-inertial (relativistic) induction equation and, finally, consider the Newtonian limit.
Because we are interested in the non-inertial equations associated with an observer locally co-moving with the fluid, it is natural to decompose the covariant Maxwell equations in the ``fluid frame'' \citep{livrev}---leading to a (fibration) formulation, commonly used in cosmology \citep{Barrow2007}.
Hence, we introduce a four-velocity $U^a$ associated with a generic observer  and decompose the Faraday tensor and the charge current as 
\begin{equation}\label{eq:Fab+current_decomp}
    F_{ab} = 2U_{[a}e_{b]} + \veps_{abc}b^c  \;, \qquad j^a = \sigma U^a + J^a \;,
\end{equation}
where
\begin{equation}\label{eq:EBdef}
    e_a =  F_{ab}U^b \;, \;\; b_a =  \frac{1}{2} \veps_{abc}F^{bc} \;, \text{ and } \veps_{abc} = \veps_{dabc}U^d \;.
\end{equation}
With these definitions we can rewrite the Maxwell equations as \citep[see e.g.][]{EllisCargese,NilsPRD12,livrev}
\begin{subequations}\label{eq:Maxwell_fibration}
\begin{equation}
    \perp^{a}_{b}\nabla_ae^b - \mu_0 \sigma =  2 W^a b_a \;,
\end{equation}
\begin{equation}
    \perp^{a}_{b}\nabla_ab^b  =  -2W^ae_a \;,
\end{equation}
\begin{multline}
    \perp_{ab} \dot e^b - \veps_{abc}\nabla^bb^c + \mu_0J_a =\\ e^b\left(\sigma_{ba} + \omega_{ba} - \frac{2}{3}\theta\perp_{ba}\right) + \veps_{abc}a^bb^c \;,
\end{multline}
\begin{multline}
    \perp_{ab}\dot b^b + \veps_{abc}\nabla^be^c =\\
    b^b \left(\sigma_{ba} + \omega_{ba} - \frac{2}{3}\theta\perp_{ba} \right) - \veps_{abc}a^be^c\;,
\end{multline}
\end{subequations}
where dots stand for co-moving time derivatives $U^a \nabla_a$ and $\perp_{ab} = g_{ab} + U_a U_b$ is the projection orthogonal to the observer four-velocity ($g_{ab}$ is the spacetime metric). 
In \cref{eq:Maxwell_fibration}, the terms on the left-hand side should be familiar, while those on the right-hand side are associated with gradients of the observer four-velocity\footnote{The shear, vorticity and expansion are defined as the four dimensional version of \cref{eq:velGradDec}---that is with additional projections to ensure they are flow orthogonal---while the observer four-acceleration is $a^b = U^a \nabla_a U^b$ and $W^a = \frac{1}{2} \veps^{abc}\om_{\change{bc}}$.}.
As such, they vanish identically for an inertial observer, and hence do not appear in most textbook discussions.

To derive the induction equation, we follow the usual logic \citep[see][for a recent relativistic discussion]{dMHD31} and ``massage'' the Faraday equation
\begin{multline}
    \underbrace{\perp_{ab}\dot b^b - b^b \left(\sigma_{ba} + \omega_{ba} - \frac{2}{3}\theta\perp_{ba} \right)}_{\sim b/T} \\
    + \underbrace{\veps_{abc}\left( \nabla^be^c  + a^be^c\right)}_{\sim e/L}  = 0 \;,
\end{multline}
to see that $e \sim L b / T \sim V b$, with $L$ and $T$  typical length- and time-scales and $V$ the associated velocity\footnote{In order to avoid confusion, let us state clearly that the scales $L,\, T$ (and the associated velocity $V$) in this argument are, in principle, not related to the ones introduced above $l_r,\, t_r$ and $v_r$. The aim here is to reproduce the usual argument for dropping the displacement current---and derive the induction equation---not to argue in which sense we can consider perturbations rapidly evolving on a non-stationary (but slowly evolving) background. When we perturb the induction equation later on, however, we are in practice assuming the displacement current to be small compared to perturbations.}.
As long as the electric and magnetic fields are slowly evolving, a similar dimensional analysis then leads us to neglecting terms involving the electric field (i.e. the displacement current) in the Ampère law, so that
\begin{equation}\label{eq:noninertial_Ampere_pre}
    J^a = \frac{1}{\mu_0} \left(\veps_{abc}\nabla^bb^c + \veps_{abc}a^bb^c\right) \;.
\end{equation}
As the Ampère law has now been demoted to a constraint on the charge current \citep[see][]{dMHD31}, we need to introduce a closure relation for the electric field. 
In the ideal case, this can be obtained by looking at the electric field measured by an observer locally co-moving with the fluid.
In fact, because the local fluid four-velocity $u^a$ is linked to the generic observer $U^a$ via
\begin{equation}
    u^a = W\left(U^a + v^a\right)\;, \quad U^a v_a = 0 \;, \quad W = \left(1- v^a v_a\right)^{-1/2}\;,
\end{equation}
where $v^b$ is the spatial fluid velocity as measured by $U^a$, the electric field measured by the fluid is (cf. \cref{eq:EBdef})
\begin{equation}
    F_{ab}u^b = W\left[e_a + \veps_{abc}v^b b^c + U_a (v^b b_b)\right]\;.
\end{equation}
In a perfect conductor, where charges easily flow, one would expect the electric field to ``short out'' as the matter becomes locally charge neutral, so that 
\begin{equation}\label{eq:ideal_Ohm}
    e_a + \veps_{abc}v^bb^c = 0 \;.
\end{equation}
With this constraint, we can derive the induction equation from Faraday's law.
To do so, note that 
\begin{multline}\label{eq:3LCcontraction_fibr}
    \veps_{abc}\veps^{cde} = U^fU_g\veps_{fabc}\veps^{gdec} = -3!U^fU_g\delta^{[g}_f\delta^{d}_a\delta^{e]}_b \\ =
    \left(\delta^d_a\delta^e_b - \delta^e_a\delta^{\change{d}}_{\change{b}}\right)- \left(\parallelsum^d_a\delta^e_b - \parallelsum^e_a\delta^d_b\right)-\left(\delta^d_a\parallelsum^e_b-\delta^e_a\parallelsum^{\change{d}}_{\change{b}}\right) \;,
\end{multline}
where we introduced the parallel projection $\parallelsum ^a_b = -U^a U_b$. 
When this is contracted with a spatial tensor (with respect to $U^a$) the last two terms in \cref{eq:3LCcontraction_fibr} can be dropped. 
It follows that 
\begin{equation}
    \veps_{abc}\veps^{cde}\left(a^b v_db_e\right) =  \left(a^b b_b\right)v_a - \left(a^b v_b\right) b_a  \;.
\end{equation}
We also need to take care of the curl of $e^a$ term in the Faraday equation. 
This can be written
\begin{multline}
    -\veps_{abc} \nabla^b\left(\veps^{cde}v_db_e\right) =\\
    \left[ -\veps_{abc} \left(\nabla^b\veps^{cde}\right)v_db_e\right] - \left[ \veps_{abc} \veps^{cde}\nabla^b \left( v_db_e\right)\right] \;,
\end{multline}
where it is convenient to consider the two terms separately. We start from the second term and, even if $U^a$ is not necessarily surface forming (i.e. has non-vanishing vorticity), we introduce a ``spatial'' covariant derivative $D$ in the usual way (projecting each index in the sub-space orthogonal to $U^a$). 
Then, it is easy to see that 
\begin{multline}
    \veps_{abc}\veps^{cde}D^b\left(v_db_e\right) =  \veps_{abc}\veps^{cde} \left(\perp^b_f\perp^g_d\perp^h_e\right)\nabla^f \left(v_gb_h\right) \\
     = \veps_{abc}\veps^{cde} \nabla^b \left(v_db_e\right) \;,
\end{multline}
and hence
\begin{equation}
    -\veps_{abc}\veps^{cde} \nabla^b \left(v_db_e\right) = D^b\left(v_bb_a\right) -D^b\left(v_ab_b\right) \;.
\end{equation}
As for the other term, writing it as
\begin{multline}
    -\veps_{abc} \left(\nabla^b\veps^{cde}\right)v_db_e =-U^g\veps_{gabc}\veps^{fcde}\left(\nabla^bU_f\right) v_db_e \\
    =-U^g\delta^{[f}_g \delta^d_a\delta^{e]}_b\,g_{fh}\left(-U^ba^h + \om^{bh} + \sigma^{bh} + \frac{1}{3}\theta\perp^{bh}\right)v_db_e \;,
\end{multline}
we see that---given the anti-symmetrization---it vanishes identically.

In summary, the (relativistic) induction equation according to a generic observer\footnote{The worldlines of the generic observer $U^a$ constitute a fibration of the spacetime, hence we may call this the ideal induction equation in the fibration framework, as opposed to the corresponding 3+1 form derived by, for example, \citet{dMHD31}.} can be written as 
\begin{multline}\label{eq:RelIndNI}
    \perp^{ab}\dot b_b  + D_b(v^b b^a) - D_b (v^a b^b)=\\  \left(\sigma^{ab} - \omega^{ab} - \frac{2}{3}\theta\perp^{ab}\right)b_b + v^a (a_b b^b) - b^a (a_b v^b) \;.
\end{multline}
The terms on the left should be familiar, while those on the right vanish for an inertial observer. 

Next, let us show how the derived equation further simplifies in the Newtonian limit.
On dimensional grounds, we observe that the last two terms on the right-hand side of \cref{eq:RelIndNI} contain an extra factor of $1/c^2$ (with respect to the others, where $c$ is the speed of light), and will as a result be negligible in the non-relativistic limit ($c^2\to \infty$). 
Similarly, let us also consider the absence of monopoles constraint. 
From \cref{eq:Maxwell_fibration} and \cref{eq:ideal_Ohm} we immediately obtain
\begin{equation}
    \perp^a_b \nabla_ab^b = 2 W^a \veps_{abc} v^b b^c \;,
\end{equation}
and we observe the term on the left hand side is $\sim b/L$ while that on the right is $\sim b L/T^2$.
Dimensional consistency implies the term on the right-hand side contains an extra factor of $1/c^2$ and should be neglected in the Newtonian limit.
In essence, non-inertial effects do not affect the absence of monopoles constraint at the Newtonian level. 
When it comes to the Lorentz force, we expect it not to change at the Newtonian level, but let us nonetheless check this for consistency. 
The Lorentz four-force can be written
\begin{multline}
     -j_bF^{ba} = - \left(\sigma U_b + J_b \right)\left(U^b e^a + U^a e^b + \veps^{bacd}U_c b_d\right) \\
     = -U^a \left(J_b\veps^{bcd}v_cb_d\right) + \veps^{abc}\left( J_b-\sigma v_b \right)b_c \;,
\end{multline}
where we used the ideal magneto-hydrodynamics relation \eqref{eq:ideal_Ohm} in the second step.
The Lorentz three force corresponds to the second term, where the charge density is measured by the observer, hence does not (in general) vanish.
However, if we insist on the local charge density to be zero---consistently with \eqref{eq:ideal_Ohm}---then we have (cf. \cref{eq:Fab+current_decomp})
\begin{equation}
    -U^a j_a = W(\sigma - v_a J^a) = 0 \Longrightarrow J_b - \sigma v_b = \left(g^a_b - v^a v_b\right)J_a \;.
\end{equation}
Re-inserting the factor of $1/c^2$ we see that the second term is negligible with respect to the first.
As also the second term in  \cref{eq:noninertial_Ampere_pre} is negligible in the Newtonian limit, we see that the Lorentz force in the Euler equation is unchanged (as expected). 

\subsection{The local frame of an observer}\label{subsec:localframe}

Having derived the relativistic induction equation according to a generic (non-inertial) observer and recalling that  we are interested in a local analysis, we now make contact with the concept of local frame associated with an observer \citep{GourghoulonSR,GravitationMTW}. 
Given an observer worldline with tangent $U^a$, the local frame is constructed by considering three spatial unit vectors that complete $U^a$ to an orthonormal basis on the tangent space at any given point. 
These unit vectors (including the observer four-velocity)---the components of which are indicated by hats---are then transported along the worldline according to 
\begin{equation}\label{eq:frameTransport}
    U^{\hat c}\nabla_{\hat c} e_{\hat a} = \Om^{\hat b}_{\;\hat a} e_{\hat b} \;,\quad \Om_{\hat a \hat b} = U_{\hat a}a_{\hat b}- a_{\hat a}U_{\hat b} -\eps_{\hat a\hat b\hat c\hat d} U^{\hat c} W^{\hat d}\;,
\end{equation}
where $a^{\hat a}$ is the four-acceleration of $U^{\hat a}$ (an intrinsic property of the worldline) and $W^{\hat d}$ is the \emph{arbitrary} four-rotation of the local frame.
Focusing on the first term in the relativistic induction equation \eqref{eq:RelIndNI}, and using \cref{eq:frameTransport},
\begin{equation}
    \dot b_{\hat b} = \left(U^{\hat c}\nabla_{\hat c} b\right)_{\hat b} = U^{\hat c} \partial_{\hat c} b_{\hat b} + \left(a^{\hat c}b_{\hat c}\right)U_{\hat b} + \veps_{\hat b \hat e\hat c}W^{\hat e} b^{\hat c}\;.
\end{equation}
The second term vanishes due to the orthogonal projection, while\footnote{Identifying the vorticity of the fibration observer with the four-rotation of the chosen local frame.}
\begin{equation}\label{eq:Wdropsout}
    \perp^{\hat a \hat b}\dot b_{\hat b} + \om^{\hat a\hat b}b_{\hat b} = \perp^{\hat a \hat b} \left(U^{\hat c}\partial_{\hat c} b_{\hat b}\right)\;.
\end{equation}
In practice, the term involving  the four-rotation of the frame drops out of the induction equation. We also note that, because we are now considering the non-inertial equations in the local frame of a single observer, there is no shear or expansion (associated with the fibration of spacetime induced by the observer). Given this,  the induction equation in the Newtonian limit simplifies to 
\begin{equation}
    \partial_t B^{\hat i} + \nabla_{\hat j}\left(v^{\hat j}B^{\hat i} - v^{\hat i}B^{\hat j}\right) = 0 \;.
\end{equation}
At the Newtonian level then, the induction equation in the local frame of a \emph{generic} observer retains the same form as for an inertial one.
This is similar to the case of the Lorentz force (entering the Euler equation) and the Ampère law. However, this is only true in the Newtonian case. Additional terms involving the four-acceleration of the observer will appear at the special relativistic level so some care will be required in order to extend our results in that direction. 

When it comes to the non-inertial terms in the Euler equations, these are obviously well known: we have to account for fictitious acceleration. 
We refer to \citet{GourghoulonSR} for a rigorous derivation in special relativity, showing how additional terms involving the observer four-acceleration also enter the relativistic expressions.
Let us also stress that working with a rotating or non-rotating local frame is entirely a matter of choice \citep{GravitationMTW}. 
At the local Newtonian level, we can always get rid of the non-inertial terms associated with the frame rotation and effectively work with the inertial equations. 

We conclude this section by noting that, as previously anticipated, some kind of filtering operation is key to separate between background and fluctuations in a highly dynamical environment. 
Postponing a discussion of this to \cref{sec:Conclusions}, let us simply note at this point that the notion of local frame discussed here is closely linked to the covariant filtering procedure discussed in \citet{fibrLES}.

\section{Going back to hydrodynamics}\label{sec:Back2Hydro}

As briefly hinted at in \cref{sec:WKBlocalbox}, the magneto-rotational instability relies on the \emph{hydrodynamic} stability of axisymmetric modes. The generic instability problem is more involved. 
If we relax the symmetry assumptions on the background, we first of all need to consider the fact that hydrodynamic shear flows tend to be unstable \citep{DrazinReid}. That is, we expect to find instabilities appearing already at the hydrodynamic level. Clearly, such instabilities would be affected by a magnetic field but not caused by it in the first place. This is an important distiction seeing as the magneto-rotational instability is \emph{due to} the presence of the  magnetic field.

With this observation in mind, let us first consider the fluid problem. 
This is important for two reasons: first, it will allow us to get a better grasp on the magnetic field impact on the instability. Second, it will allow us to make contact with the Rayleigh criterion (and ultimately the magneto-rotational instability). 
As the fluid problem is much simpler than the magneto-fluid one, we will study the case where both shear and vorticity gradients are retained, and also discuss the impact of shear viscosity---either of microphysical origin or due to filtering\footnote{Any filtering operation will introduce additional \emph{residual} terms every time it acts upon a non-linear term \citep[][]{LesieurLES,lesbook,mcdonough,SchmidtLES}. These terms are akin to (but not quite the same as) dissipative terms \citep{fibrLES}, and are meant to capture transport to/from unresolved scales.}.
Shear viscosity is introduced in the usual way \citep[see][]{LandauFLuidMechanics}, and the shear viscosity coefficient $\eta$ will be considered constant, consistent with the local analysis.

Because the calculation is a bit tedious and we want to focus on the implications for the physics, the derivation of the relevant fluid dispersion relation(s) is provided in  \cref{app:HydroDispRel}. There we also express the coefficients of the resulting characteristic polynomial(s) in terms of scalars built from the background quantities. This allows us to keep the discussion as general as possible, without having to refer to a specific background configuration. 
It is, however, worth stressing that, in many situations of interest the relevant dynamics is either sub- or supersonic. Given this, for these problems it is worth considering models that filter out modes that are either faster or slower than the sound waves. 
This can be done starting from a fully compressible dispersion relation and taking  either of two  limits: either we assume the speed of sound to be very large, in which case the model becomes sound-proof \citep[we point to][for more details]{Vasil_2013}, or very small. 
In the following, we typically work in the sound-proof limit, noting that the MRI is typically discussed within the so-called Boussinesq approximation \citep[][]{BoussinesqBarletta}, thus removing fast magneto-sonic waves\footnote{They do, however, retain perturbations in the fluid pressure in the Euler equation as they consider a non-barotropic equation of state and the impact of stratification.} \citep[][]{BalbusHawley1}.

Starting from the continuity equation, perturbing it and introducing the plane-wave expansion we readily obtain
\begin{equation}\label{eq:MHDpertCont}
    \partial_t \delta \rho + \delta\rho \nabla_i v^i + v^i \nabla_i \delta\rho + \rho \nabla_i \delta v^i  = 0  \Longrightarrow -i \omega \delta\rho + i \rho k_i \delta v^i =0 \;, 
\end{equation}
where $\om$ and $k_i$ are defined as in \cref{sec:WKBlocalbox}.
Note that we set $v^i =0$ as we evaluate the relation at the centre of the local box, and assume that  the background expansion rate $\nabla_i v^i$ can be neglected. Similarly, we write the perturbed Euler equation including a shear-viscous term as
\begin{multline}\label{eq:HydroEulerVisc}
     \partial_t \delta  v_i + \delta v^j \nabla_j v_i + \frac{1}{\rho} \nabla_i \delta P - \delta \left(\eta \nabla^j \tau_{ji}\right) = 0 \\
     \Longrightarrow -i\omega \delta v_i + i \frac{c_s^2 }{\rho} k_i \delta \rho + \sigma_{ij}\delta v^j + \eps_{ijk} W^j \delta v^k - \delta \left(\eta \nabla^j \tau_{ji}\right)=0 \;,
\end{multline}
where $\tau_{ji}$ is the rate-of-strain tensor and $W^i = 1/2 \eps^{ijk}\om_{jk}$. 
Working this out we have retained gradients in the background flow only, used the velocity gradient decomposition (\cref{subsec:vel_gradients}), introduced the adiabatic speed of sound $c_s^2 = \partial P / \partial \rho$, and considered the gravitational potential to be externally sourced (hence neglecting its perturbations). 

As a first example, consider a background with negligible vorticity, set the shear viscosity to zero and take the sound-proof limit. 
Then consider the case $\text{det}(\gvec\sigma) =0$, and first of all look for modes such that $\sigma_{ij}k^j =0$. 
Recalling that, as discussed in \cref{subsec:vel_gradients}, a mainly two dimensional flow with negligible vorticity is characterized by having a shear matrix with vanishing determinant---that is $\text{det}(\gvec\sigma) \sim \O(\eps)$---and noting that we can choose the orientation of the local axes in such a way that the background flow is, say, along the $\hat x,\,\hat y$ directions, we can always consider the determinant to be zero. 
This means that there always exists a wave-vector living in the eigen-space corresponding to the zero eigenvalue\footnote{\change{We note that these modes correspond to those propagating in the $z$-direction in terms of the adapted Cartesian coordinates introduced in \cref{subsec:vel_gradients}. We also stress that the same condition is satisfied by the fastest growing unstable modes, which also propagate vertically, although with respect to the ``global'' cylindrical system.}}. 
Taking these steps, we end up with\footnote{Note that the same dispersion relation applies in the opposite limit where $c_s$ is small.} (cf. \cref{eq:HydroShearOnly})  
\begin{equation}\label{eq:HydroFastestModes}
    \om^2 = -\frac{1}{2}\text{Tr}(\gvec \sigma^2)  \Longrightarrow \om = \pm i \sqrt{\frac{1}{2}\text{Tr}(\gvec \sigma^2) }\;.
\end{equation}
These modes are non-propagating, and half of them are unstable with a growth rate independent of the wave-vector. 
Next, we consider wave-numbers such that\footnote{\change{Using the adapted coordinates of \cref{subsec:vel_gradients} it is possible to see that these wave-vectors have non-zero $x$ and $y$ components, the ratio between the two being equal to that of the independent entries of the shear matrix. More importantly, it can be seen that the same condition is satisfied by the axisymmetric modes usually considered in the derivation of the magneto-rotational instability.}} $\sigma_{ij} k^i k^j = 0$ (but $\sigma_{ij}k^j \neq 0 $), noting that such modes will always exist---even in the more general case (considered below) where $\text{det}(\gvec\sigma)\neq 0$. 
It follows that for such modes
\begin{equation}\label{eq:HydroNext2Fastest}
    \om^2 = -\frac{1}{6}\text{Tr}(\gvec\sigma^2)\Longrightarrow \om = \pm i \sqrt{\frac{1}{6}\text{Tr}(\gvec \sigma^2) } \;.
\end{equation}
These modes are also non-propagating, and half of them are unstable with a (constant) growth rate about a factor of 2 smaller than in the previous case. 
As the dispersion relation is quadratic (in the sound-proof limit), we can explicitly solve it and confirm the expectation (and well known fact) that shearing flows are generically unstable.
\change{Given the role played by $\text{Tr}(\gvec\sigma^2)$ here and in what follows, let us briefly comment on how one can intuitively see why this makes sense. 
First, let us note that the same term enters the growth of entropy in the generalized second law equation.
Second, we also note that the same term when evaluated on the usual background for the magneto-rotational instability would lead to $\text{Tr}(\boldsymbol{\sigma}^2)= 4 (\text{d}\Omega / \text{dln}R)^2$. As such a term enters the usual instability dispersion relation and criterion, it is not surprising it plays a pivotal role also for the present discussion. }   

Let us now build on this and discuss how vorticity and shear viscosity impact on the generic instability of shearing flows. 
First consider the case where the background has negligible vorticity but non-vanishing shear viscosity, observing that the corresponding modes in a homogeneous background are stable provided $\eta >0$. If viscosity is  of microphysical origin, then $\eta >0$ follows from the second law of thermodynamics \citep[see][]{LandauFLuidMechanics,livrev}. 
If the viscosity is instead due to filtering, a positive value of $\eta$ corresponds to an eddy-type model where energy is cascading to smaller/unresolved scales\footnote{\change{We also note that it is in principle possible to have $\eta<0$ when this is not of micro-physical origin but an effective viscosity instead. Negative values of $\eta$ would correspond, in the simplest eddy-viscosity-type models to net energy source at the resolved scales, and hence an inverse cascade from the unresolved ones (see, e.g. section VIB in \cite{fibrLES}).}} \citep{LesieurLES,lesbook,mcdonough,SchmidtLES}. 
Going back to the case with both shear and viscosity (in the sound-proof limit), the dispersion relation of modes such that $\sigma_{ij} k^j =0$ is (cf. \cref{eq:HydroShearVisc})
\begin{equation}
    \om^2 +i \eta k^2 \om - \frac{1}{4}\left[ \eta^2 (k^2)^2 - 2 \text{Tr}(\gvec\sigma^2) \right] =0 \;,
\end{equation}
where $k = |\vec k| $. Assuming $\eta >0$, stability corresponds to
\begin{equation}
    \eta^2 (k^2)^2 - 2 \text{Tr}(\gvec\sigma^2)  > 0 \;.
\end{equation}
In essence, comparing this to \cref{eq:HydroFastestModes} we see that viscosity tends to stabilize shear-unstable modes, with a larger impact at smaller scales. This makes intuitive sense.
Next, consider modes such that $\sigma_{ij} k^i  k^j =0$, which are solutions of
\begin{equation}\label{eq:HydroNext2FastestEta}
     \om^2 + i\eta k^2 \om - \frac{1}{12}\left[ 3\eta^2 (k^2)^2 - 2 \text{Tr}(\gvec\sigma^2) \right] =0 \;.
\end{equation}
As before, these modes---to be compared with their counterparts in \cref{eq:HydroNext2Fastest}---are also stable (assuming $\eta>0$) provided the last term in \cref{eq:HydroNext2FastestEta} is negative. This will be true when the wave-number is sufficiently large, and we have verified that the same trend is true for generic wave-vectors.
In essence, we learn (as one may have expected) that shear viscosity generically slows the growth rate of unstable shear modes, and stabilises modes with small enough wavelengths. 

Turning to the case where the background has non-negligible vorticity and shear, and taking the sound-proof limit as before, we first observe that the fastest growing modes encountered before, namely those characterized by $\sigma_{ij} k^j =0$, are not guaranteed to exist anymore, as the determinant of the shear matrix cannot  in general be assumed to be negligible (see \cref{subsec:vel_gradients}). 
Should these modes exist, though, their dispersion relation would be (cf. \cref{eq:S+Whydro})
\begin{multline}
    \om^2 = -\frac{1}{2}\text{Tr}(\gvec\sigma^2) + (\hat k\cdot \vec W)^2 \\
    \Longrightarrow \om = \pm i \sqrt{\frac{1}{2}\text{Tr}(\gvec\sigma^2) - (\hat k\cdot \vec W)^2 }\;, \quad \hat k = \vec k / k
\end{multline}
and we see, comparing this to \cref{eq:HydroFastestModes}, that vorticity tends to stabilize them. 
We also observe that---in contrast to shear viscosity---vorticity affects all such modes by reducing their growth rate in a way that does not depend on their wave-number (although the direction of propagation is important). 
Next---and also because the modes we just looked at may not exist---we consider modes such that $\sigma_{ij} k^i  k^j = 0$, with dispersion relation
\begin{equation}
    \om^2 = -\frac{1}{6}\text{Tr}(\gvec\sigma^2) + (\hat k\cdot \vec W)^2\;.
\end{equation}
Comparing this to \cref{eq:HydroNext2Fastest}, we observe again that vorticity tends to stabilize such modes in a way that does not depend on the wave-number.
We have verified that the same trend is also true for generic wave-vectors. 
As a final point, it is easy to verify that the case with only background vorticity is generally stable (not only in the sound-proof limit). 

In summary, a sheared background flow is generically unstable already at the hydrodynamic level, which is a well-known fact.
However, we have considered the impact that shear viscosity and/or vorticity have on the instability of the possible hydrodynamic modes. The results show that shear viscosity tends to weaken the instability  in general, with larger effects for larger wave-numbers. Meanwhile,   vorticity has a stabilizing effect which does not depend on the wave-number. 
Finally, let us also point to \cref{app:RayleighCloser} where we show that the general dispersion relation derived in \cref{app:HydroDispRel} (discussed here) is shown to encompass the classic Rayleigh stability criterion. 

\section{Magneto-shear instability in the local frame}\label{sec:MSinstaLocal}

Having explored the hydrodynamic case, let us  perturb the corresponding MHD equations and study the impact of the magnetic field on the generic shear instabilities we encountered.
We consider a barotropic equation of state and retain gradients in the background velocity only, as we want to focus on the magneto-shear nature of the instability \citep[cf.][]{BalbusHawley2,shibataNR}. 
The continuity equation is obviously unchanged, while the perturbed Euler equation becomes
\begin{multline}\label{eq:MHDpertEuler}
    \partial_t \delta  v_i + \delta v^j \nabla_j v_i + \frac{1}{\rho}\nabla_i \delta P + \frac{1}{\mu_0\rho} \left[ B_j \nabla_i \delta B^j - B^j \nabla_j \delta B_i\right] = 0 \\ \Longrightarrow -i\omega \delta v_i + i \frac{c_s^2 }{\rho} k_i \delta \rho + \frac{i}{\mu_0\rho} \left[ (B_j \delta B^j)k_i - (B^j k_j)\delta B_i \right] \\
    + \sigma_{ij}\delta v^j + \eps_{ijk} W^j \delta v^k =0 \;.
\end{multline}
Finally, the perturbed induction equation is
\begin{multline}\label{eq:MHDpertInd}
    \partial_t \delta B^i + B^i \nabla_j \delta v^j - B^j \nabla_j \delta v^i - \delta B^j \nabla_j v^i + \delta B^i \nabla_j v^j  =0 \\ \Longrightarrow -i\omega \delta B^i + i B^i(k_j \delta v^j ) - i (B^j k_j)\delta v^i \\
    - \sigma^{ij}\delta B_j -  \eps^{ijk}W_k \delta B_j + \frac{2}{3}\theta \delta B^i= 0 \;.
\end{multline}
We will now discuss the linearized system that follows from these equations, as before focussing on the results and providing more detailed steps in \cref{app:MHDdispRel}. 

Let us first recap the mode analysis for the homogeneous case. 
In order to derive the fully compressible dispersion relation, we first re-scale the magnetic field as 
\begin{equation}\label{eq:RescaleB}
    \vec v_A \doteq \frac{\vec B}{\sqrt{\mu_0\rho}} \;,\quad   \delta \vec v_A \doteq \frac{\delta\vec B}{\sqrt{\mu_0\rho}} \;.
\end{equation}
Then, computing the dispersion relation from the linearized equations above (neglecting background shear and vorticity) we obtain (cf. \cref{eqapp:FCMHDwaves})
\begin{equation}\label{eq:FCMHDwaves}
    -\om \left(\om^2 - \left(\vec v_A \cdot \vec k\right)^2 \right)\left[ \om^4 - \left(v_A^2 + c_s^2 \right)k ^2 \om^2 + c_s^2  k ^2 \left(\vec v_A \cdot \vec k\right)^2 \right] =0 \;,
\end{equation}
where the roots of the second factor correspond to Alfv\'{e}n waves, while those of the quartic polynomial in square brackets describe fast and slow magneto-sonic waves. 
Before moving on to discuss the impact of shear and vorticity, let us briefly note what happens to such modes when we take the sound-proof limit---in which the speed of sound is large.
From \cref{eq:FCMHDwaves} we see that in the first case fast magneto-sonic waves are filtered out, while the slow ones reduce to Alfv\'{e}n waves. 
In the opposite limit---when disturbances are much faster than the sound waves---the dispersion relation describes Alfv\'{e}n waves and the low-$c_s$ limit of fast magneto-sonic waves. 
This limit corresponds to ignoring fluid pressure perturbations while retaining variations in the magnetic pressure. 

Let us now consider the impact of a non-negligible background shear. 
Again, we focus on the results, with the details of the derivation provided in \cref{subapp:ShearCase}. 
We start from the fully compressible dispersion relation and consider the sound-proof limit, leading to the dispersion relation \eqref{eq:ShearedCoeffSP}. 
As in the hydrodynamic case considered earlier, we first consider the case $\text{det}(\gvec\sigma) =0$, and look for modes such that $\sigma^{ij}k_j =0$. 
It is then easy to see that the general dispersion relation  \eqref{eq:ShearedCoeffSP} simplifies to (ignoring a trivial root)
\begin{equation}
    \left[\om^2 - \left(\frac{1}{2}\text{Tr}(\gvec\sigma^2) - (\vec v_A \cdot\vec k)^2\right)\right]^2 = 0\;.
\end{equation}
Comparing to the corresponding hydrodynamic modes in \cref{eq:HydroFastestModes}, we immediately see that the magnetic field tends to have a stabilizing effect (provided it is not orthogonal to the wave-vector, in which case it has no effect whatsoever).

\begin{figure*}
\centering
\includegraphics[width=1\textwidth]{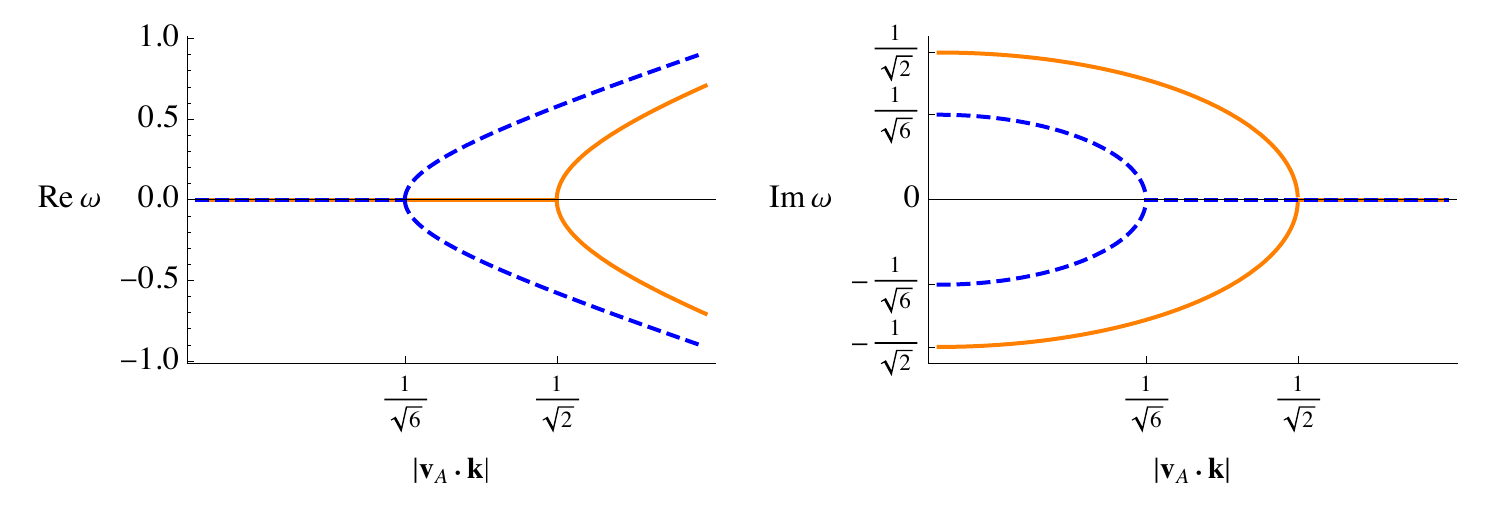}
\caption{
Real and imaginary part of the solutions of \cref{eq:N2FMHDdr}, with the frequency and $|\vec v_A\cdot \vec k|$ in units of $\sqrt{\text{Tr}(\gvec \sigma^2)}$.
The solutions plotted correspond to the fastest growing modes evolving on top of an MHD sheared background. 
We see that the magnetic field has a stabilizing effect, as the growth rates are reduced with respect to those of the corresponding hydrodynamic modes. 
The stabilizing effect is all the more pronounced the more the wave-vector is aligned with the magnetic field lines, and is switched off for modes propagating in the directions perpendicular to the magnetic field lines.
In particular, modes corresponding to sufficiently large values of $|\vec v_A\cdot \vec k|$ are turned stable. 
}
\label{fig:N2FinstaMHD}
\end{figure*}

Next, we take (again, as before) $\text{det}(\gvec\sigma) =0$ and consider modes such that $\sigma_{ij}k^ik^j =0$ (but $\sigma_{ij}k^j \neq 0$).
The relevant dispersion relation can then be written (making use of \cref{eq:N2Ftrick})
\begin{equation}\label{eq:N2FMHDdr}
    \om^4 + b_2 \om^2 + b_4 = 0\;,
\end{equation}
with 
\begin{subequations}
\begin{equation}
    b_2 = \frac{2}{3}\text{Tr}(\gvec\sigma^2) - 2 (\vec v_A \cdot\vec k)^2\;,
\end{equation}
and 
\begin{equation}
    b_4 = \frac{1}{12}\text{Tr}(\gvec\sigma^2)^2 - \frac{2}{3}\text{Tr}(\gvec\sigma^2)(\vec v_A \cdot\vec k)^2 + (\vec v_A \cdot\vec k)^4\;.
\end{equation}
\end{subequations}
The stabilizing effect of the magnetic field is evident from \cref{fig:N2FinstaMHD}, where both the frequency and $|\vec v_A\cdot\vec k|$ are plotted in units of $\sqrt{\text{Tr}(\gvec\sigma^2)}$.
The key point here is that, while the background shear is required for the instability (the vanishing-shear modes are stable Alfvèn waves in the sound-proof limit), the magnetic field is not the main driver.
This is evident from the results as the imaginary part of the unstable modes remains finite in the limit $\vec v_A \to 0$, and the limiting value coincides with the hydrodynamic result (from the previous section).
This observation, possibly unexpected at first sight, deserves a thorough discussion, and we will return to this issue in  \cref{subsec:InstaVsMRI}. 
Before we expand on that aspect, let us stress that the results make intuitive sense.
The magnetic field impacts on the instability in that it breaks the hydrodynamic isotropy and dampens the growth of unstable modes propagating along magnetic field lines. 
This also suggests that shear-instability driven turbulence is isotropic in the hydrodynamic case but inherently anisotropic for magnetized flows, consistently with the overall picture \citep[see e.g.][]{SchekochihinRev,Beresnyak2019Rev}. 
Before moving on, it is also worth noting 
that the background velocity profile considered by Balbus \& Hawley  \citep{BalbusHawley1,BalbusHawleyRev} is characterized by having a shear matrix with vanishing determinant (and expansion rate), and also that for axisymmetric modes $\sigma_{ij}k^ik^j = 0$, while  the fastest growing MRI modes propagate vertically, with $\sigma_{ij}k^j =0$.

As a final comment before we make contact with the usual MRI and the Rayleigh criterion, we have also considered the case with non-negligible background vorticity only. The details are discussed in \cref{subapp:VorticityCase}, but the crucial result is that magnetized flows are generically stable in this case. As this feature is unchanged from the corresponding fluid case, it is  reasonable to expect that the same trend we discussed for the purely hydrodynamical case will also apply to the magnetized case with both shear and vorticity: Vorticity tends to stabilize shear-unstable modes in a manner independent of the wave number (\change{although the orientation of the wave-vector with respect to the vorticity is expected to have an impact.}).

\section{Concluding Remarks: The instability in perspective}\label{sec:Conclusions}

We set out with the intention of discussing the magneto-rotational instability in a general background, relaxing the symmetry constraints associated with the standard analysis and possibly deriving an instability ``criterion'' relevant for (highly) dynamical environments and nonlinear simulations. 
\change{In fact, the usual criterion is often used to compute a  quality factor for the instability, and hence measure how well the mechanism ``could be resolved'' in a given simulation.}
However, having set up the analysis (and the required tools) in an arguably sensible way, we arrived at results which were not in line with the ``naïve'' expectations. Given this, it makes sense to comment  on the implications. Moreover, we need to  highlight an important ``missing ingredient'' in the discussion; the need to involve some suitable filtering operation to make the discussion sensible in the first place.  We will deal with each of these questions in turn, starting with the implications of our results for the MRI. 

\subsection{The  MRI vs the Rayleigh criterion}\label{subsec:InstaVsMRI}

A key  aspect of  the MRI is that adding a weak magnetic field on top of a hydrodynamically stable shearing flow changes the nature of the problem and makes it unstable. 
In discussing this problem, however, it is often ``forgotten'' that the relevant hydrodynamic stability criterion \citep{Rayleigh1917} guarantees stability \underline{only for} axisymmetric modes (cf. the discussion in \cref{app:RayleighCloser}). Adding a magnetic field renders such modes unstable---technically, the non-axisymmetric ones are not \citep{BalbusHawley4}. 
Thus it is clear that the MRI is relevant only in situations where we can think of axisymmetric modes as being  ``preferred'' in some sense. An immediate example of this is an accretion disk, which involves a  globally axisymmetric background for the perturbations. This then immediately tells us that applying the results to the dynamical context of neutron star mergers is a much more subtle endeavour.
In fact, this exercise is problematic from the outset.

To back up this claim, we show in \cref{app:MRIlocal} that we can reproduce the MRI perturbation equations and dispersion relation through the local frame construction. 
However, for the specific MRI calculation there exists a  preferred local frame: the co-rotating frame. 
This local frame is set up considering an observer that is co-rotating with the fluid along some orbit, and the coordinate axes  rotate in such a way that one of them always points in the radial direction of a global cylindrical coordinate system. Another coordinate axis  always points in the azimuthal direction. 
This local frame is  ``preferred'' as the axes are (by construction) tied to those of the most natural global coordinate system. In a sense, we could set up different local co-rotating observers and construct the global axes by stitching together the local ones. 
In the case of a general and truly local analysis, however, this additional piece of information is not available. 

Moreover, we show in \cref{app:RayleighCloser} how one may set up (for the circular and axisymmetric background flow) a local frame that is ``co-moving but not co-rotating'' with the fluid. 
In doing so, we derived the corresponding dispersion relation, confirmed that the result is consistent with the general formulae (cf. \cref{app:HydroDispRel}), and showed how we can recover the usual Rayleigh criterion (and hence also the MRI criterion) as long as we perform the conversion to the relevant co-rotating frame frequency.

These arguments clarify the sense in which the MRI (and similarly the Rayleigh stability) is a ``global instability analyzed with local tools''. The local analysis needs to be ``augmented'' by pieces of information that cannot be truly local. 
The upshot of this is that, in a merger-like scenario (where assumptions regarding the global properties of the flow are debatable) we should probably not expect the standard instability criteria to provide a faithful indication/diagnostic of what is actually going on. The standard argument will apply, but only if there is a meaningful sense of (Rayleigh stable) flow on a scale larger than that at which the plane-wave analysis is carried out. This would make a discussion much more difficult for any given numerical simulation, but so be it.

\subsection{The missing ingredient: Filtering}

Throughout the discussion we have focussed on the analytical development, sweeping issues associated with actual numerical data ``under the carpet''. 
The key issue here is that we ignored the question of how one would, in practice, construct the background suitable for the perturbation analysis given nonlinear simulation dynamics. 
\change{If the numerical data shows some degree of axisymmetry, then the split could be achieved via an azimuthal averaging (and the usual criteria will apply). If this is not the case, however,}
\remove{In words, the answer is easy:} we need to apply some suitable filtering operation to remove small scale fluctuations from a gradually varying ``background''.
\change{With such a filtering procedure at hand we could start from real numerical data and perform the split into background plus fluctuations. The background data could then be used as numerical input in the dispersion relations we have discussed to assess whether or not the instability is active and how rapidly it grows. }
In a nonlinear setting this split is (obviously) not guaranteed to make sense. Suppose that the instability we are trying to uncover acts on some characteristic scale $L$, say. Then we need a background that varies on a larger scale than this, otherwise the notion of a shear flow that becomes unstable due to smaller scale waves makes no sense.  This argument relies on an explicit filtering step, separating the instability scale $L$ from the variation of the background. The construction of such a filter should be possible, at least in principle, in many situations (see, for example, \citet{fibrLES}).
Of course, the scale separation may not apply in actual problems of interest. 
\change{In addition, the analysis in this work (like any linear stability analysis) relies on the expansion detailed in \cref{sec:WKBlocalbox} and the neglect of non-linear fluctuations. The intention is to describe the early stages of the development of the instability. Crucially, because the split into background and fluctuations depends on the filtering procedure, the requirement that fluctuations are indeed small will further constrain the procedure (e.g. the filter width).}

Further complicating the discussion is the unavoidable implicit filtering associated with the finite numerical resolution. We know from the large body of work on turbulence simulations that sub-grid dynamics may play an important role in a robust description of the dynamics. This typically involves a suitable large-eddy scheme to represent the subgrid dynamics. Hence, the analysis involves elements of choice (effectively, the closure relations).
Crucially, the effective field theory that is/should be simulated is not that of the ideal theory. 
All current models---both the ones discussed in \citet{carrasco} and \citet{radice2} as well as the covariant scheme of \citet{fibrLES}---modify the principal part of the equations of motion. 
Therefore the analysis of the model ``that is actually solved'' is fundamentally changed, even when the closure terms are small.
In essence, an instability analysis of numerical simulation data needs to consider the impact of an effective viscosity/resistivity. 
Given the presently available tools, we do not have a particularly good handle on this issue. 
We are forced to conclude that we also need to make progress on the development of robust large-eddy models before we can make a sensible attempt to demonstrate the presence of the MRI in a highly dynamical environment.

\section*{Acknowledgements}
NA and IH gratefully acknowledge support from Science and Technology Facility Council (STFC) via grant numbers ST/R00045X/1 and ST/V000551/1. 

\section*{Data availability}

Supplementary data underlying this article are available at the following links: \href{https://doi.org/10.5281/zenodo.7612469}{https://doi.org/10.5281/zenodo.7612469} and \href{https://doi.org/10.5281/zenodo.7612479}{https://doi.org/10.5281/zenodo.7612479}.
\bibliographystyle{mnras}
\bibliography{MRI_LES.bib}

\begin{appendix}
    \section{Formulating the MRI in the local frame}\label{app:MRIlocal}
In this appendix we derive the MRI using the local frame construction discussed in the main text. 
We consider the circular velocity profile assumed in \citet{BalbusHawley1}, $\vec v = v^{\hat \varp} \hat \varp$ with $ v^{\hat \varp} = \Om(R) R$ where we use cylindrical coordinates and an orthonormal basis on the ``tangent space'' (as usual).
We then distinguish between indices with a ``hat'' corresponding to the orthonormal basis, and those without that correspond to the coordinate basis. 
We then pick an orbit at some radial distance $R_0$ and choose an observer that is co-rotating with angular frequency identical to that of the background flow at $R_0$, that is $\vec v_{obs} = \Omega_0 R \hat\varp$ where $\Om_0 = \Om(R_0)$.
The observer is then accelerated with acceleration $\vec a = - \Om_0^2 R \hat R$, and the velocity of the fluid with respect to such an observer then is $\vec v' = (\Om - \Om_0)R\hat\varp$.
We then set up the axes of the observer's local frame so that one is pointing in the radial direction ($\hat e_1$), one is pointing in the azimuthal direction ($\hat e_2$) and the third one is aligned with the rotation axis ($\hat e_3$). 
Introducing coordinates associated with this observer, we can then write the background fluid velocity as 
\begin{equation}
    \vec v' = \frac{\text{d}\Om}{\text{dln}R}\bigg|_{R_0} x' \hat e_2  + \O(x'^2)\;.
\end{equation}
We have neglected terms of order $\O\left(x'^2\right)$ as we will only need the velocity and its gradients evaluated at the origin of the frame---so that such terms will not enter the perturbation equations anyway.
Computing the gradients we then obtain 
\begin{equation}\label{eq:CRgrad}
    \partial'_iv'_j = \begin{pmatrix}
    0 & s_0 & 0 \\
    0 & 0 & 0 \\
    0 & 0 & 0
    \end{pmatrix} \;,\qquad s_0 = \frac{\text{d}\Om}{\text{dln}R}\bigg|_{R_0} \;.
\end{equation}
As the local frame of the observer is rotating with angular velocity $\Om_0 \hat e_3$, we need to include the Coriolis force into the perturbation equations. 
We then write the perturbed Euler and continuity equations (dropping the primes for clarity, and retaining only gradients in the background velocity) as
\begin{subequations}
\begin{equation}
    \partial_t \delta \rho + \rho \partial_i \delta v^i = 0 \;,
\end{equation}
\begin{multline}
    \partial_t \delta v_i + 2\Om_0 \eps_{i3k} \delta v^k + \delta v^j \partial_j v_i + \frac{c_s^2}{\rho}\partial_i \delta \rho \\
    + \frac{1}{\mu_0\rho} \left[ B_j \partial_i \delta B^j - B^j \partial_j \delta B_i\right] = 0 \;,
\end{multline}
\end{subequations}
and, introducing a WKB plane-wave expansion,
\begin{subequations}\label{eq:Euler+contMRI}
\begin{equation}
    -i \om \delta \rho + i \rho k_i \delta v^i = 0 \;,
\end{equation}
\begin{multline}
    -i \om \delta v_i + 2\Om_0 \eps_{i3k}\delta v^k + s_0 \delta_{i2}\delta v^1 + i \frac{c_s^2}{\rho}k_i \delta \rho \\
    + \frac{i}{\mu_0 \rho}\left[B_jk_i \delta B^j - B^j k_j \delta B_i\right] =0 \;,
\end{multline}
\end{subequations}
Next, focus on the induction equation. As we have discussed above, the induction equation in the co-rotating frame retains the inertial form. 
We then have 
\begin{equation}
    \delta \left[\partial_j \left(v^j B^i -v^i B^j\right) \right]= \delta v^j \partial_j B^i + v^j \partial_j \delta B^i - B^j \partial_j \delta v^i - \delta B^j \partial_j v^i \;,
\end{equation}
where we made use of the no-monopoles constraint, the vanishing expansion  of the background flow and got rid of the divergence of the perturbed velocity consistently with the Boussinesq approximation. 
Introducing the WKB plane-wave expansion and evaluating the background quantities at the origin of the local frame we then end up with 
\begin{equation}\label{eq:InductionMRI}
    - i \om \delta B^i - i B^j k_j \delta v^i - \delta B^1 s_0 \delta ^{i2} = 0 \;.
\end{equation}
In \cref{eq:Euler+contMRI,eq:InductionMRI} we recognize the terms entering the perturbation equations in \cite{BalbusHawley1} (with the exception of gradients in the background pressure that we are neglecting here). 
We also note that we do not need to formally neglect terms of the form $B/ R$ as these terms do not appear in the explicit local frame construction. 
We conclude by noting that, at the special relativistic level, a uniformly rotating observer and the co-rotating one are not the same as the latter is also accelerated \citep[see][ch. 13]{GourghoulonSR}. However, this difference is irrelevant at the level of the Newtonian perturbation equations since i) pseudo-acceleration terms drop out of the perturbed Euler equation ii) non-inertial terms in the induction equation involving the four-acceleration are negligible in the Newtonian limit.

\subsection{Another look at the non-inertial MHD equations}
Before we move on to take a closer look at the Rayleigh criterion, let us show how the terms involving the local frame rotation drop out of the induction equation.
Even though we have already argued this happens in general (cf. \cref{eq:Wdropsout}), we here prove this for the specific case of a co-rotating observer. 
We do so as this allows us to appreciate better why the cancellation comes about.
We will use a notation that is common in general relativity, that is the notion of spin coefficients associated with a non-coordinate basis \citep{CarrollST}.
The covariant derivative of a tensor $T^{\hat a}_{\hat b}$ is
\begin{equation}
    \nabla_a T^{\hat a}_{\hat b} = \partial_a T^{\hat a}_{\hat b} + \om_{a\;\hat c}^{\;\hat a}T^{\hat c}_{\hat b} - \om_{a\;\hat b}^{\;\hat c}T^{\hat a}_{\hat c} \;,
\end{equation}
with
\begin{equation}\label{eq:spinCoeffDef}
    \om_{a\;\hat c}^{\;\hat b} = e^{\hat c}_{d}\,e^e_{\hat b}\,\Gamma^{d}_{ae} - e^d_{\hat b}\partial_a e^{\hat c}_{d} \;,
\end{equation}
where $\Gamma^{a}_{bc}$ are the connection coefficients associated with the coordinates chosen while $e^{\hat a}_b$ is the matrix connecting the coordinate basis to the orthonormal one. 
We also note here that for a local frame we identify $u^a \om_{a\; \hat b}^{\;\hat a} = \Om^{\hat a}_{\;\hat b}$ introduced in \cref{eq:frameTransport}.
We then introduce (Born) coordinates associated with a uniformly rotating observer (axes suitably oriented so that the angular velocity is $\Om_0 \hat z$)
\begin{equation}\label{eq:BornCoord}
    t' = t \;,\quad z' = z \;,\quad x' = R \text{cos}(\Om_0 t + \varphi) \;,\quad y' = R \text{sin}(\Om_0 t + \varp)\;,
\end{equation}
where primed coordinates are Cartesian (i.e. non-rotating). Computing the spin coefficients (starting from a flat metric) we then obtain 
\begin{equation}\label{eq:BornSC}
     \om_{\varphi\;\hat \varp}^{\;\hat R} = -1 \;,\quad \om_{\varphi\;\hat R}^{\;\hat \varp} = +1 \;,\quad
     \om_{t\;\hat \varp}^{\;\hat R} = -\Om_0 \;,\quad
     \om_{t\;\hat R}^{\;\hat \varp} = \Om_0 \;,
\end{equation}
showing that, as the coordinates ``mix space and time'' we need to introduce a covariant derivative in the time-direction as well.
We then write the non-inertial induction equation as (cf. \cref{eq:RelIndNI})
\begin{equation}\label{eq:CorotatingInduction}
    \nabla_tB^{\hat i} +\nabla_{\hat j} \left(v^{\hat j} B^{\hat i} - v^{\hat i} B^{\hat j} \right) + \eps^{\hat i \hat j\hat k}B_{\hat j}\Om^0_{\hat k}  = 0\;.
\end{equation}
It is then easy to verify, by means of \cref{eq:BornSC} that
\begin{equation}
    \nabla_t B^{\hat i } + \eps^{\hat i \hat j\hat k}  B_{\hat j}\Om^0_{\hat k}=  \partial_t  B^{\hat i} \;,
\end{equation}
thus confirming the result in \cref{eq:Wdropsout} and the use of the inertial induction equations.
We stress that the co-rotating frame rotation vector is $\Omega_0 \hat e_3$, and is the same (by construction) as the vorticity of the observer. This is why we see the same cancellation as
in \cref{eq:Wdropsout}, where we assumed they are equal.

\section{A closer look at the Rayleigh stability criterion}\label{app:RayleighCloser}

The key point of the magneto-rotational instability is that the circular velocity background is stable against axisymmetric hydrodynamic perturbations, while adding a (however weak) magnetic field changes completely the nature of the system and makes it unstable to such perturbations. We now revisit the Rayleigh criterion in order to highlight the key role played by the co-rotating observer in deriving the criterion. These important aspects have to be kept in mind when looking at the general results derived below (cf. \cref{app:HydroDispRel,app:MHDdispRel}) and discussed in \cref{sec:MSinstaLocal,sec:Back2Hydro}.

Starting from \cref{eq:Euler+contMRI}, and ignoring terms associated with the magnetic field, we write the coefficient matrix  (ordering the perturbed quantities as $\{\delta \rho/\rho, \delta v^1, \delta v^2, \delta v^3 \}$)
\begin{equation}\label{eq:CMRayleighCR}
    \begin{pmatrix}
        -\om & k_1 & k_2 & k_3 \\
        c_s^2 k_1 & - \om & 2 i \Om_0 & 0 \\
        c_s^2 k_2 & - i \frac{\kappa^2}{2\Om_0} & - \om & 0 \\
        c_s^2 k_3 & 0 & 0 & - \om 
    \end{pmatrix} \;, 
\end{equation}
where 
\begin{equation}
     \frac{\kappa^2}{2\Om_0} = 2 \Om_0 + s_0 = 2 \Om_0 + \frac{\text{d}\Om}{\text{dln}R}\bigg|_{R_0}\;,
\end{equation}
and the dispersion relation reads 
\begin{equation}
    \om^4 - \left(c_s^2 k^2 - \kappa^2 \right)\om^2 + i c_s^2 s_0 k_1k_2\om + c_s^2 \kappa^2 (k_3)^2 = 0 \;.
\end{equation}
Taking the sound-proof limit we then end up with 
\begin{equation}
    k^2 \om^2 - i s_0 k_1 k_2 \om - \kappa^2 k_3^2 = 0\;.
\end{equation}
From this we easily see that, if we assume axisymmetric perturbations, namely $k_2=0$, we obtain the usual Rayleigh stability criterion, that is $\kappa^2 > 0$ \citep{Rayleigh1917}. We stress that, as is well-known, the criterion does not guarantee that non-axisymmetric modes are stable. In fact, rewriting the dispersion relation in terms of $\Delta = - i\om$ and using the Routh-Hurwitz criterion \citep{korn2013mathematical} we find that, on top of the Rayleigh criterion we would also need  
\begin{equation}
    s_0 k_1 k_2 \le 0 \;.
\end{equation}
We also note that, the story changes if we take the opposite limit, in which case the Rayleigh criterion is sufficient to guarantee stability of non-axisymmetric perturbations as well. This would also be the case had we assumed incompressibility from the start. 

Having discussed the usual Rayleigh criterion using the co-rotating observer, we now re-work it using an observer that is orbiting with the fluid at a given orbital distance $R_0$ but whose (local frame) axes are non rotating. 
We do this for two reasons. 
First, it will allow for a direct comparison with the general results in \cref{app:HydroDispRel}.
Second, we have argued that choosing to work with a rotating or non-rotating observer is, in general, just a matter of taste.
We then pick up an orbit $R_0$ as before, and choose the observer to be co-orbiting with the background flow at the specific orbit
\begin{equation}
    \vec v_{obs} =  - \Om_0 y_0 \hat x + \Om_0 x_0 \hat y\;,
\end{equation}
where we used global Cartesian coordinates and $x_0(t) = R_0 \text{cos}(\Om_0 t), y_0(t) = R_0 \text{sin}(\Om_0 t)$ describe the worldline of the observer (the origin of the axes is suitably chosen so that $z_0(t) = 0$).
The background fluid velocity is then 
\begin{equation}
    \vec v = - \Om x \hat y + \Om y \hat x \;,\qquad \Om = \Om (\sqrt{x^2 + y^2}) \;,
\end{equation}
so that, considering the relative velocity $\vec v' = \vec v - \vec v_{obs}$ and expanding around $(x_0, y_0)$ we obtain
\begin{multline}
    \vec v' = - \left[s_0 \frac{x_0 y_0}{x_0^2 + y_0^2} x' + \left(\Om_0 + s_0 \frac{y_0^2}{x_0^2 +y_0^2}\right)y'\right]\hat x \\+ \left[ \left(\Om_0 + s_0 \frac{x_0^2}{x_0^2 +y_0^2}\right)x' + s_0 \frac{x_0 y_0}{x_0^2 + y_0^2} x' \right]\hat y \;,
\end{multline}
where $x' = x-x_0,\, y' = y - y_0$.
We can now choose a local region around a specific point $(x_0,y_0,z_0)$ on the orbit and choose to re-orient the axes by a constant rotation so that the observer velocity is moving only in the $y-$direction. 
We then set up the local frame in such a way that the local axes are non-rotating and oriented like the global cartesian ones. 
We can therefore write the gradients as
\begin{equation}\label{eq:NRgrad}
    \partial'_iv'_j = \begin{pmatrix}
    0 & \Om_0 + s_0 & 0 \\
    -\Om_0 & 0 & 0 \\
    0 & 0 & 0
    \end{pmatrix} \;,
\end{equation}
and the coefficient matrix of the linearized Euler plus continuity system is (cf. \cref{eq:Euler+contMRI} and ignore both magnetic field terms and the Coriolis force as the axis are non-rotating)
\begin{equation}\label{eq:CMRayleighNR}
    \begin{pmatrix}
        -\om & k_1 & k_2 & k_3 \\
        c_s^2 k_1 & - \om & i \Om_0 & 0 \\
        c_s^2 k_2 & - i (\Om_0 + s_0) & - \om & 0 \\
        c_s^2 k_3 & 0 & 0 & - \om 
    \end{pmatrix} \;.
\end{equation}
We can then compute the dispersion relation to find 
\begin{multline}
    \om^4 - \left[c_s^2 k^2 + \Om_0(\Om_0 + s_0) \right]\om^2 + i c_s^2 k_1k_2 s_0\, \om \\
    + c_s^2 \Om_0(\Om_0 + s_0) (k_3)^2 = 0 \;,
\end{multline}
and observe this is consistent with the general dispersion relation in \cref{eq:S+Whydro} when restricted to the shear and vorticity associated with \cref{eq:NRgrad}.
However, this is not quite the dispersion relation we obtained above. 
The reason for this is that the two local observers we have considered measure different frequencies, as the axes of the co-rotating observer rotate with angular velocity $\Om_0 \hat e_3$ with respect to the other. 
To show why this is the resolution to the apparent conflict, let us consider once again the Born coordinates (cf. \cref{eq:BornCoord,eq:BornSC}). Given any vector $a^{\hat i}$ we have 
\begin{equation}
    \nabla_t a^{\hat i} = \partial_t \hat a^i + \Om_0 \eps^{\hat i \hat 3 \hat k} a_{\hat k}\;.
\end{equation}
This relation, when we introduce a plane-wave WKB expansion translates to %
\begin{equation}\label{eq:NRvsCRfrequencyConversion}
    - i \om_{rot} \delta a^{\hat i} = -i \om_{nr} \delta a^{\hat i}  + \Om_0 \eps^{\hat i \hat 3 \hat k} \delta a_{\hat k}\;,
\end{equation}
where $\om_{rot}$ is the frequency measured by the co-rotating observer, while $\om_{nr}$ is the frequency measured by an observer that has the same worldline but uses non-rotating axes. 
Specifying \cref{eq:NRvsCRfrequencyConversion} to the perturbed velocity (noting that it would not apply to the continuity equation as the density is a scalar), and noting that the frequency in \cref{eq:CMRayleighNR} corresponds to $\om_{nr}$, we can reconcile the results obtained from \cref{eq:CMRayleighNR} with those from \cref{eq:CMRayleighCR}. 
We also note here that the same logic applies when we consider magnetized flows. 
That is, if we work with the inertial induction equation and compute the background velocity gradients as in \cref{eq:NRgrad}, we also need to take into account the relation in \cref{eq:NRvsCRfrequencyConversion} for magnetic field disturbances to get back to \cref{eq:InductionMRI} and the MRI dispersion relation.

\section{Hydrodynamic dispersion relations}\label{app:HydroDispRel}
In this appendix we provide more details of the results discussed in \cref{sec:Back2Hydro} of the main text. 
In order to derive the dispersion relation and study the effects of a sheared background, it is convenient to choose a basis that is adapted to it.
Because the shear is a symmetric trace-free matrix there exists a basis such that  
\begin{equation}\label{eq:ShearBasis}
    \sigma_{ij} = \text{diag} \left(\sigma_1, \sigma_2, - (\sigma_1 + \sigma_2)\right) \;.
\end{equation}
We will make use of this basis to write down the coefficients matrix of the linearized system.
Before doing so, however, it is reasonable to wonder whether this change of basis has any impact on the perturbation equations. We are, in fact always free to choose a basis in the tangent space that is not associated with the coordinates chosen, but this (in general) introduces additional terms in the covariant derivative. Let us spell out why this is not the case here. Working with a non-coordinate basis, we need to account for spin-coefficients---defined as in \cref{eq:spinCoeffDef}---when a derivative acts on vectors and tensors. 
The first term in \cref{eq:spinCoeffDef} then vanishes as we are here working with a non-rotating Cartesian frame (so that the Christoffel symbols vanish), while the second term is in general non vanishing, and accounts for the fact that the change of basis (in the tangent space) needed to diagonalize the shear matrix may change from point to point. In the context of this analysis, however, we are looking at scales smaller than those over which background quantities vary. 
In essence, also this second term vanishes as the shear matrix is (by construction) constant over the local region of fluid we are zooming on.

Working in the shear adapted basis (cf. \cref{eq:ShearBasis}), we write the coefficient matrix of the linearized system as (cf. \cref{eq:HydroEulerVisc})
\begin{equation}
    \begin{pmatrix}
    -\om & \rho k_1 & \rho k_2 & \rho k_3 \\
    \frac{c_s^2}{\rho}k_1 & - \om - i \sigma_1 - i \eta L_1 & i W^3 - \frac{i}{6} \eta k_1 k_2 & - i W^2 - \frac{i}{6}\eta k_1 k_3\\
    \frac{c_s^2}{\rho}k_2 &- i W^3 - \frac{i}{6}\eta k_2 k_1 &  - \om - i \sigma_2 - i \eta L_2 & i W^1 - \frac{i}{6}\eta k_2 k_3 \\
    \frac{c_s^2}{\rho}k_3 &  i W^2 - \frac{i}{6}\eta k_3 k_1 & - i W^1 - \frac{i}{6}\eta  k_3 k_2 &  - \om + i \sigma_3- i \eta L_3
    \end{pmatrix}\;,
\end{equation}
where $\sigma_3 = \sigma_1 + \sigma_2$,
$L_1 = \frac{2}{3} k_1 ^2 + \frac{1}{2}k_2 ^2 + \frac{1}{2} k_3^2$ and $L_2,\,L_3$ are similarly defined. 

The dispersion relation is computed taking the determinant of this matrix and equating it to zero.
In order to keep the discussion as general as possible (i.e. without having to refer to a specific background configuration) we will decompose the coefficients of the characteristic polynomial in terms of scalars built from background quantities. 
In the simplest cases this can be done “by eye”, but the procedure can easily become quite messy. 
The logic is nonetheless simple: we group the different terms in each coefficients according to the power of the various background quantities, for example we group all the terms quadratic in the shear and wave-vector components.
We then build all the possible scalars that are quadratic in the shear and wave-vector, and look for the correct linear combination of them. 
This logic can be easily implemented on a computer algebra program like Mathematica\footnote{See \href{https://doi.org/10.5281/zenodo.7612469}{https://doi.org/10.5281/zenodo.7612469} for more details.}.
We now discuss the dispersion relations obtained by retaining only shear terms, both shear and viscous terms, and lastly shear and vorticity
terms.

Before doing so, we first observe that the coefficients will involve scalars constructed from the shear matrix only.
As with any $3\cross3$ matrix, the shear matrix $\gvec \sigma$ has three invariants 
\begin{equation}
    I_1 = \text{Tr}(\gvec \sigma) \;, \quad I_2 = \frac{1}{2}\left[ \text{Tr}(\gvec \sigma^2) - \left( \text{Tr}(\gvec \sigma)\right)^2\right]\;, \quad I_3 = \text{det}(\gvec \sigma)\;,
\end{equation}
related via the Cayley-Hamilton theorem as
\begin{equation}
    \gvec \sigma^3 - I_1 \gvec \sigma^2 + I_2 \gvec \sigma - I_3 \mathbb{I} = 0 \;,
\end{equation}
where $\mathbb{I}$ is the $3\cross3$ identity matrix.
Because the shear matrix is trace-free, we will write the coefficients in terms of $\sfrac{1}{2}\text{Tr}(\gvec\sigma^2)$ and $\text{det}(\gvec\sigma)$.
It is also useful to note that there will always exist modes such that $\sigma_{ij}k^i k^j =0$. In the convenient shear basis, these are characterized by $k^1 = k^2 = k^3$ if the determinant is not vanishing (that is $s_1 \neq s_2$), and $k^1 = k^2$ when it does.
It follows that for such modes
\begin{equation}\label{eq:N2Ftrick}
    -\frac{1}{2}\text{Tr}(\gvec\sigma^2) + \sigma^2_{ij}\hat k^i \hat k^j  = - \frac{1}{6}\text{Tr}(\gvec\sigma^2)  \;,\quad \hat k = \vec k / |\vec k| = \vec k/ k \;.
\end{equation}

The resulting dispersion relation for the case with negligible vorticity and viscosity is then 
\begin{equation}\label{eq:HydroShearOnly}
    \om^4 + a_2 \om^2 + a_1 \om + a_0 = 0
\end{equation}
with 
\begin{subequations}
\begin{align}
    a_2 & = -c_s^2 k^2 + \frac{1}{2}\text{Tr}(\gvec \sigma^2) \;,\\
    a_1 &= i \left[c_s^2 \sigma_{ij}k^ik^j - \text{det}(\sigma)\right] \;,\\
    a_0 & = c_s^2 \left[\sigma^2_{ij}k^ik^j -\frac{1}{2}\text{Tr}(\gvec \sigma^2)  k^2\right]\;,
\end{align}
\end{subequations}
whose solutions are discussed in the main text (cf. \cref{sec:Back2Hydro}).

Turning to the case with non-vanishing shear viscosity, the dispersion relation is then 
\begin{equation}\label{eq:HydroShearVisc}
    \om^4 + a_3 \om^3 + a_2 \om^2 + a_1 \om + a_0 = 0\;,
\end{equation}
with 
\begin{subequations}
\begin{align}
    a_3 &= \frac{5}{3}i \eta k^2 \;, \\
    a_2 &= -c_s^2 k^2 + \frac{1}{2}\text{Tr}(\gvec\sigma^2) - \frac{1}{12}\eta \left[11 \eta (k^2)^2 -2 \sigma_{ij} k^i  k^j\right] \;, \\
    a_0 &= c_s^2 \left[\sigma^2_{ij} k^i  k^j - \frac{1}{2}\text{Tr}(\gvec\sigma^2) k^2 - \frac{1}{2}\eta k^2 \sigma_{ij} k^i  k^j + \frac{1}{4} \eta^2 (k^2)^3\right]\;,
\end{align}
and 
\begin{multline}
     a_1 = i c_s^2 \left[\sigma_{ij} k^i  k^j - \eta (k^2)^2\right]\\ 
     + i \Big\{ -\frac{1}{6} \left[\sigma^2_{ij} k^i k^j -2 \text{Tr}(\gvec \sigma^2)k^2 \right] \\+ \frac{1}{12}\eta^2 k^2 (\sigma_{ij} k^i  k^j) 
     - \frac{1}{6}\eta^3 (k^2)^3 - \text{det}(\gvec \sigma)\Big\}\;.
\end{multline}
\end{subequations}
We can sanity check this dispersion relation by considering the homogeneous background limit.
It is then immediate to see that, in the sound-proof limit, this would be stable provided $\eta >0$. 
It turns out that this condition guarantees stability even outside of the sound-proof limit, as can be verified by means of the Routh-Hurwitz criterion \citep{korn2013mathematical}.
As discussed in the main text, this is the case both when shear viscosity is of micro-physical origin (due to the second law of thermodynamics) but also when this is an effective viscosity due to filtering provided this models a net energy transfer to smaller unresolved scales---which is intuitive.

Finally, the dispersion relation in the case where the background has non-negligible vorticity and shear is
\begin{equation}\label{eq:S+Whydro}
    \om^4 + a_2 \om^2 + a_1 \om + a_0 = 0 \;,
\end{equation}
with 
\begin{subequations}
\begin{align}
    a_2 &= - c_s^2 k^2 - W^2 + \frac{1}{2}\text{Tr}(\gvec\sigma^2) \;,\\
    a_1 &= i \left[c_s^2 \sigma_{ij}k^i k^j - \text{det}(\gvec\sigma) - \sigma_{ij}W^i W^j\right] \;,\\
    a_0 &= c_s^2 \left[  \sigma^2_{ij}k^i k^j - \frac{1}{2}\text{Tr}(\gvec\sigma^2)k^2 + (\vec k \cdot \vec W)^2\right]\;.
\end{align}
\end{subequations}

\section{MHD dispersion relations}\label{app:MHDdispRel}
In this appendix we provide more details on the results discussed in \cref{sec:MSinstaLocal} of the main text.
In order to keep the presentation tidy, we will discuss separately the homogeneous background case (cf. \cref{subapp:HomogeneousBG}), the sheared case (cf. \cref{subapp:ShearCase}) and the case with non-negligible background vorticity (cf. \cref{subapp:VorticityCase}). 

\subsection{Homogeneous background}\label{subapp:HomogeneousBG}
In order to derive the fully compressible dispersion relation for the homogeneous case, we first re-scale the magnetic field as in \cref{eq:RescaleB}, and introduce a convenient basis $\{\hat v_A,\, \hat q,\,\hat s \}$ where $\hat v_A =\vec v_A /|\vec v_A|$ while $\hat q,\hat s$ complete it to an orthonormal basis.
For instance, assuming $\vec v_A$ is not aligned with $\vec k$ we can construct it as
\begin{equation}
    \vec q = \vec k - \left(\vec k \cdot \hat v_A\right)\hat v_A \,,\qquad \hat q = \frac{\vec q}{|\vec q|} \,, \qquad \hat s = \hat v_A \cross\hat q \;,
\end{equation}
so that\footnote{If the wave-vector is along the background magnetic field we just have to set $k^q =0$ in the following.} 
\begin{equation}
    \vec k = k^{v_A} \hat {v_A} + k^q \hat q  \;.
\end{equation}
The coefficient matrix of the linearized system can then be written as (cf. \cref{eq:MHDpertCont,eq:MHDpertEuler,eq:MHDpertInd} and ignore background vorticity and shear)
\begin{equation}
    \vec M = \begin{pmatrix}
    \vec A & \vec C \\
    \vec C^\top & \vec D
    \end{pmatrix}
\end{equation}
with 
\begin{subequations}
\begin{equation}
    \vec A = \begin{pmatrix}
        -\om & \rho k_{v_A} & \rho k_q & 0 \\
        \frac{c_s^2}{\rho} k_{v_A} & - \om & 0 & 0 \\
        \frac{c_s^2}{\rho} k_q & 0 & - \om & 0 \\
        0 & 0 & 0 & -\om 
    \end{pmatrix} \;, 
\end{equation}
\begin{equation}
    \vec C = \begin{pmatrix}
        0 & 0 & 0 \\
        0 & 0 & 0 \\
        b k_q & - b k_{v_A} & 0 \\
        0 & 0 & - b k_{v_A}
    \end{pmatrix} \;,
\end{equation}
and
\begin{equation}
    \vec D = \begin{pmatrix}
        -\om & 0 & 0 \\
        0 & - \om & 0 \\
        0 & 0 & -\om
    \end{pmatrix} \;.
\end{equation}
\end{subequations}
As $\vec D$ is clearly invertible, we can reduce $\vec M$ into factors via the Schur complement
\begin{multline}
    \begin{pmatrix}
    \vec A & \vec C \\
    \vec C^\top & \vec D
    \end{pmatrix} \\
    = 
    \begin{pmatrix}
    \mathbf{I}_4 &  \vec C \vec D^{-1} \\
    \vec 0_{4 \times 3} & \mathbf{I}_3
    \end{pmatrix} 
    \begin{pmatrix}
    \vec A - \vec C \vec D^{-1} \vec C ^\top & \vec 0_{4\times 3} \\
    \vec 0_{3\times 4} & \vec D
    \end{pmatrix}
    \begin{pmatrix}
    \mathbf{I}_4 &  \vec 0_{4 \times 3} \\
    \vec D^{-1} \vec C ^\top & \mathbf{I}_3
    \end{pmatrix}\;,
\end{multline}
and then compute the determinant as
\begin{equation}
    \text{det}(\vec M) = \text{det}(\vec D)\, \text{det}(\vec A - \vec C \vec D^{-1} \vec C ^\top) \;.
\end{equation}
The resulting dispersion relation is 
\begin{equation}\label{eqapp:FCMHDwaves}
    -\om \left(\om^2 - \left(\vec v_A \cdot \vec k\right)^2 \right)\left[ \om^4 - \left(v_A^2 + c_s^2 \right) k ^2 \om^2 + c_s^2  k ^2 \left(\vec v_A \cdot \vec k\right)^2 \right] =0 \;.
\end{equation}

\subsection{Sheared Background}\label{subapp:ShearCase}
Let us now consider the case where the background vorticity is negligible while shear terms are not.  
Re-scaling the magnetic field as in \cref{eq:RescaleB} and decomposing \cref{eq:MHDpertCont,eq:MHDpertEuler,eq:MHDpertInd} (ignoring vorticity terms) as well as $\delta\vec v $ and $\delta \vec v_A$ in the shear-adapted basis, the coefficient matrix of the linearized system of equations reads 
\begin{equation}\label{eq:ShearedMatrix}
    \vec M = \begin{pmatrix}
        \vec A & \vec C \\
        \vec C^\top & \vec D
    \end{pmatrix} \;,
\end{equation}
where
\begin{subequations}
\begin{equation}
    \vec A = \begin{pmatrix}
        - \om & \rho k_1 & \rho k_2 & \rho k_3 \\
        \frac{c_s^2}{\rho} k_1 & - \om -i \sigma_1 & 0 & 0 \\
        \frac{c_s^2}{\rho} k_2 & 0 & -\om -i \sigma_2 & 0 \\
        \frac{c_s^2}{\rho} k_3 & 0 & 0 & - \om +i (\sigma_1 + \sigma_2) 
    \end{pmatrix} \;,
\end{equation}
\begin{equation}
    \vec D = \begin{pmatrix}
        -\om + i \sigma_1 & 0 & 0 \\
        0 & -\om + i \sigma_2 & 0 \\
        0 & 0 & -\om - i (\sigma_1 + \sigma_2) 
    \end{pmatrix} \;,
\end{equation}
\renewcommand\arraystretch{1.4}
\begin{equation}
    \vec C = \begin{pmatrix}
        0 & 0 & 0 \\
        I_1 & v_A^2k_1  & v_A^3 k_1 \\
        v_A^1k_2 & I_2 & v_A^3 k_2 \\
        v_A^1k_3 & v_A^2k_3 & I_3  
    \end{pmatrix} \;,
\end{equation}
\end{subequations}
while
\begin{equation}
    I_1 = v_A^1 k_1 - \left(\vec v_A\cdot\vec k\right) \;,
\end{equation}
and $I_2,I_3$ are defined similarly. 

In a similar fashion as for the hydrodynamic case considered above, we will decompose the coefficients of the characteristic polynomial in terms of scalars built from background quantities. 
As we might have expected, the resulting dispersion relation is a complicated seventh-degree polynomial\footnote{See \href{https://doi.org/10.5281/zenodo.7612479}{https://doi.org/10.5281/zenodo.7612479} for the details of the scalar decomposition.} (and we sanity-checked it reduces to the homogeneous case when we set to vanish the shear terms).
In order to learn something useful out of it, we then consider the sound-proof limit and retain only terms proportional to the speed of sound.
We end up with the following dispersion relation
\begin{equation}\label{eq:ShearedCoeffSP}
    a_5 \om^5 + a_4 \om^4 + a_3\om^3 + a_2 \om^2 + a_1 \om + a_0 = 0 \;,
\end{equation}
with
\begin{subequations}
\begin{multline}
    a_0 = - i \bigg\{\text{det}(\gvec \sigma)\Big[\sigma^2_{ij}k^ik^j - \frac{1}{2}\text{Tr}(\gvec{\sigma}^2)\Big] \\
    + (\vec v_A\cdot\vec k)^2 \Big[\text{det}(\gvec \sigma)k^2 -\frac{1}{2}(\sigma_{ij}k^ik^j)\text{Tr}(\gvec \sigma^2)  \Big] + (\vec v_A\cdot\vec k)^4\sigma_{ij}k^ik^j\bigg\} \;,
\end{multline}
\begin{multline}
    a_1 =  \bigg\{(\vec v_A\cdot\vec k)^4 k^2 + \left(\vec v_A \cdot \vec k\right)^2\left[\sigma^2_{ij}k^ik^j - \text{Tr}(\gvec\sigma^2)k^2\right] +\\
    \text{det}(\gvec\sigma)\left(\sigma_{ij}k^ik^j\right) + \frac{1}{2}\text{Tr}(\gvec\sigma^2)\left[\frac{1}{2}\text{Tr}(\gvec\sigma^2)k^2 - \sigma^2_{ij}k^ik^j \right]\bigg\} \;,
\end{multline}
\begin{equation}
    a_2 = i \Bigg\{ -\frac{1}{2}(\sigma_{ij}k^ik^j)\text{Tr}(\gvec\sigma^2) + \text{det}(\gvec \sigma)k^2 +2 (\vec v_A\cdot\vec k)^2 (\sigma_{ij}k^ik^j)\Bigg\}\;,  
\end{equation}
\begin{equation}
    a_3 =  \left[\text{Tr}(\gvec\sigma^2)k^2 -2 (\vec v_A\cdot\vec k)^2k^2 - \sigma^2_{ij}k^ik^j\right]\;,
\end{equation}
\begin{equation}
    a_4 = -i (\sigma_{ij}k^ik^j)\;,
\end{equation}
\begin{equation}
    a_5 =  k^2 \;.
\end{equation}
\end{subequations}

\subsection{Background with vorticity}\label{subapp:VorticityCase}
We now turn to the case with non-negligible background vorticity (but negligible shear). We re-scale the magnetic field as in \cref{eq:RescaleB} and introduce a convenient basis $\{\hat W,\, \hat q,\,\hat s \}$, where $\hat W =\vec W /|\vec W|$ while $\hat q,\hat s$ complete it to an orthonormal basis. 
For instance, assuming $\vec v_A$ is not aligned with $\vec W$ we can construct it as
\begin{equation}
    \vec q = \vec v_A - \left(\vec v_A \cdot \hat W\right)\hat W \,,\qquad \hat q = \frac{\vec q}{|\vec q|} \,, \qquad \hat s = \hat W \cross\hat q \;,
\end{equation}
and the magnetic field\footnote{Note that the definition of $\hat q$ changes when the background magnetic field is aligned with the vorticity, even though in what follows we would simply have to set $v_A^q=0$.}  
\begin{equation}
    \vec v_A = v_A^W \hat W + v_A^q\hat q \;.
\end{equation}

The coefficient matrix of the linearized system then is (cf. \cref{eq:MHDpertCont,eq:MHDpertEuler,eq:MHDpertInd} and ignore shear terms)
\begin{equation}
    \vec M = \begin{pmatrix}
        \vec A & \vec C \\
        \vec C^\top & \vec D
    \end{pmatrix} \;,
\end{equation}
with 
\begin{subequations}
\renewcommand\arraystretch{1.4}
\begin{equation}
    \vec C = \begin{pmatrix}
        0 & 0 & 0 \\
        -v_A^q k^q & v_A^q k^W & 0\\
        v_A^W k^q & - v_A^ W k^W & 0 \\
        v_A^W k^s & v_A^q k^s & - \left(v_A^W k^W + v_A^q k^q\right) 
    \end{pmatrix}\;,
\end{equation}
\renewcommand\arraystretch{1.4}
\begin{equation}
    \vec A = \begin{pmatrix}
        -\om & \rho k^W & \rho k^q & \rho k^s \\
        \frac{c_s^2}{\rho}k^W & - \om & 0 & 0 \\
        \frac{c_s^2}{\rho}k^q & 0 & - \om & +i W \\
        \frac{c_s^2}{\rho}k^s & 0 & -i W & - \om 
    \end{pmatrix} \;, 
\end{equation}
and 
\begin{equation}
    \vec D = \begin{pmatrix}
        - \om & 0 & 0 \\
        0 & - \om & - i W \\
        0 & + i W & - \om
    \end{pmatrix}
\end{equation}
\end{subequations}

As for the sheared case, we compute the dispersion relation by taking the determinant of this matrix and decompose each coefficient as a sum of scalars\footnote{More details on the decomposition can be found at \href{https://doi.org/10.5281/zenodo.7612479}{https://doi.org/10.5281/zenodo.7612479}.}. 
Having sanity-checked the result by contrasting it against the homogeneous background dispersion relation, we take the sound proof limit.
The sound-proof dispersion relation can then be written as
\begin{equation}\label{eq:VortDR}
    \om^4 + b_2 \om^2 + b_4 = 0\;,
\end{equation}
with 
\begin{subequations}\label{eq:VortStabQuartic}
\begin{align}
    b_2 &= -\left[W^2 +  (\hat k \cdot\vec W)^2 + 2 (\vec v_A\cdot\vec k)^2 \right]\;, \\
    b_4 &=   \left[(\vec v_A \cdot\vec k)^2 + W^2 \right]  \left[ (\hat k \cdot \vec W)^2 + (\vec v_A\cdot\vec k)^2\right] \;.
\end{align}
\end{subequations}
 As this is a particularly simple quartic polynomial, we can study the stability of its roots analytically. 
Considering \cref{eq:VortDR} as an equation for $\om^2$ and computing the discriminant we obtain %
\begin{equation}
    \left[ W^2 - (\hat k \cdot \vec W)^2\right]^2 \ge 0
\end{equation}
so that $\om^2$-roots are real. As complex roots of a real algebraic polynomials occur in pairs of complex conjugates, complex $\om^2$-roots would correspond to an instability. 
In order to have stable roots though, we also need other conditions to be met. 
We, in fact need $b_2<0$ and $b_4>0$ to make sure that the $\om^2$-roots are real and positive, so that $\om$-roots are real as well. 
As this is evidently the case, we conclude that background vorticity---just like in the purely hydrodynamic case---does not lead to an instability. 

\end{appendix}

\end{document}